\documentclass[showpacs,preprint,amsmath,superscriptaddress,prl]{revtex4}  % alvaro modified
\usepackage[dvips]{graphicx}
\usepackage{epsfig}
\usepackage{color}
\usepackage{soul}

\begin{document} 
\title{
%Spontaneous 
Increased
extinction probability
of the Madden-Julian Oscillation 
after about 27 days
%from the RMM Index
}
\author{
\'Alvaro Corral
}
\affiliation{%
Centre de Recerca Matem\`atica,
Edifici C, Campus Bellaterra,
E-08193 Barcelona, Spain
}\affiliation{Departament de Matem\`atiques,
Facultat de Ci\`encies,
Universitat Aut\`onoma de Barcelona,
E-08193 Barcelona, Spain}
%\affiliation{Barcelona Graduate School of Mathematics, Edifici C, Campus Bellaterra, E-08193 Barcelona, Spain}
\affiliation{Complexity Science Hub Vienna,
Josefst\"adter Stra$\beta$e 39,
1080 Vienna,
Austria
}
\email{alvaro.corral@uab.es}
\author{
M\'onica Minjares
}
\affiliation{%
Centre de Recerca Matem\`atica,
Edifici C, Campus Bellaterra,
E-08193 Barcelona, Spain
}\affiliation{Departament de F\'{\i}sica,
Facultat de Ci\`encies,
Universitat Aut\`onoma de Barcelona,
E-08193 Barcelona, Spain}
\author{
Marcelo Barreiro
}
\affiliation{%
Departamento de Ciencias de la Atm\'osfera, Facultad
de Ciencias, Universidad de la Rep\'ublica, Igua 4225,
11400 Montevideo, Uruguay
}
\begin{abstract} 
The Madden-Julian oscillation (MJO) is a tropical weather system
having important influence in the tropics and beyond;
however, many of its characteristics are poorly understood, 
including their initiation and completion.
Here we define Madden-Julian events as the contiguous time periods
with an active MJO, 
and we show that both the durations and the sizes of these events are 
well described by a double power-law distribution.
Thus, small events have no characteristic scale, 
and the same for large events;
nevertheless, both types of events are separated by a characteristic duration of about 27 days (this corresponds to half a cycle, roughly). 
Thus, after 27 days, there is a sharp increase in the probability that an event gets extinct.
We find that this effect is independent of the starting and ending phases of the events, 
which seems to point 
%to be due to an internal mechanism 
to an internal mechanism of exhaustion
rather than to the effect of an external barrier.
Our results would imply an important limitation of the MJO as a driver of sub-seasonal predictability.
\end{abstract} 
% \pacs{64.60.av, 91.30.-f, 81.30.Kf,05.50.+q}

\date{\today}

\maketitle

\section{Introduction}

The Earth system presents 
%structures 
variability
across an enormous range of temporal, spatial, and energy scales
\cite{Ben_Zion_review,Bodenschatz,Williams_census,Ghil_Lucarini,Franzke}.
%e.g.,
In the case of %geophysical fluids, 
weather and climate, 
%there are processes ranging 
%there is a range going 
%there are
these go from
small turbulent structures,
%from turbulent motion, 
% vertical turbulent mixing % ghil
% turbulent eddies % franzke
in the scales of a few seconds and centimeters, to ice ages,
in tens of thousands of years at the planetary span.
%
%... citar franzke, citar ghil \& Lucarini !!.
%
%In the case of weather and climate, 
This variability has been traditionally assigned to well-defined peaks in the power spectrum of the weather-climate signals, corresponding to certain periodic or nearly-cyclic 
%%external forcings
processes 
(such as daily cycles, annual cycles, Milankovich cycles, etc.)
that dominate over a secondary background noise.
However, it is starting to be recognized that the most important part 
in the weather and climate variability
comes precisely from the ``background’’ \cite{Franzke}.
It is a remarkable fact that the variability that arises at each scale turns out to be ``similar’’ to that at other scales, so that self-similarity characterizes the weather-climate system. 
Needless to say, deep understanding of this variability is of fundamental importance to improve predictability 
and acknowledge its limitations
(see, e.g., Refs. \cite{Deluca_Moloney_Corral,Corral_Elsner}),
as well as to distinguish between natural and anthropogenic trends;
nevertheless, the self-similarity paradigm 
(also called scale invariance or scaling)
is far from having entered into the mainstream practice of Earth sciences \cite{Franzke}.

% algo de sensitivity analysis??
% “assessment of future climate changes caused by anthropogenic forcing”
%the structures that arise at each scale

Self-similarity presents many different aspects. 
In time series, 
one may find it 
in the probability density of the measured variable (the signal),
in the form of a power-law tail \cite{Yang_Franzke,Franzke}, 
which enhances the probability of extreme records
\cite{footnote_mjo1}
and implies that large values of the variable can arise 
without a characteristic scale.
Self-similarity can also 
manifest
in a power-law shape of the power spectrum, 
signaling the absence of a finite correlation time,
which implies correlation patterns 
%and clustering of extremes 
at all temporal scales,
including the clustering of extremes.
%in the absence of a finite correlation time, 
%so that the power spectrum (related to the autocorrelation %function) follows a power-law decay. 
%Further, 
%In addition 
%

%
As an alternative
to perform the statistics for individual measurements of the signal 
(such as hourly rain), 
one can define events, 
for which the signal is above 
%(or below) 
a fixed threshold for a certain period of time
(such as rain events at a fixed location).
The durations of these events can be power-law distributed also \cite{Peters_pre}, 
as well as the integral of the signal along event duration 
(referred to as the size or sometimes, roughly speaking, as the ``energy'' of the event) \cite{Corral_Gonzalez}. 
Waiting times between consecutive events have been found to be power-law distributed in some cases 
\cite{Peters_Deluca,Deluca_npg,%
%franzke2013persistent,%esta ref no la encuentra!!! OJO!!!
%yang2020power,%esta tampoco
Benzi_21}, 
although in some others self-similarity is manifested 
%not only in the form of power laws but 
not in the power-law shape of the distributions but 
in bivariate scaling laws relating waiting time and a threshold in the value of the signal or in event size \cite{Bunde,Corral_csf,Corral_fires,Morina_storms}.

Using time series defined over a spatial grid 
(constituting an evolving geophysical field), 
one may be able to construct spatio-temporal events (or ``clusters'') of activated signals, which again have been claimed to be power-law distributed 
in terms of their spatial size
\cite{Peters_dragon_kings,Traxl,Corral_Gonzalez}.
On the purely geometric side, 
fractal structures are another signature of self-similarity,
and have been found for instance in clouds \cite{Lovejoy81,Lovejoy}.

%Franzke, C. L. E. (2013). Persistent regimes and extreme events of the North Atlantic atmospheric circulation. Philosophical Transactions of the Royal Society A, 371(1991), 20110471.??
% no veo ninguna power law aqui!!

%Yang, L., Franzke, C. L. E., \& Fu, Z. (2019). Power-law behavior of hourly precipitation in intensity and dry spell durations over the US.
%International Journal of Climatology, 8, 1–16.
% aqui no veo claro que pasa en los waitings, 
% pero las intensidades dicen que tienen cola power law

%AQUI LA FRAGILIDAD DE LAS POWER LAWS??
%\cite{Corral_nuclear,Corral_Deluca,Barabasi_criticism,Voitalov_krioukov}

%Of particular interest for us 
Particularly relevant for our purposes
is the case of tropical cyclones 
(comprising hurricanes, typhoons, tropical storms, tropical depressions...).
Tropical cyclones are routinely 
%recognized 
identified
as individual spatio-temporal events,
for which the maximum sustained wind speed is recorded every 6 hours, 
from onset to dissipation.
Integration of the cube of this speed along the lifetime of the tropical cyclone
%of the tropical cyclone 
yields the so-called power dissipation index \cite{Emanuel_nature05}, 
which is a proxy of the total energy dissipated  
by the tropical cyclone
\cite{Corral_hurricanes,Corral_agu}. 
Reference \cite{Corral_hurricanes} found that the tropical-cyclone energy estimated in this way follows a power-law distribution, with nearly the same exponent for all tropical-cyclone basins,
although the most extreme events (in terms of energy)
turned out to be not power-law but exponentially-like distributed,
due to finite-size effects that break the power-law decay.
The results found by Traxl et al. \cite{Traxl} for spatio-temporal rainfall clusters are probably another side of the same scale-free phenomenon, 
but extended to extratropical systems.

This abundance of power laws and scaling in meteorology and climatology
%difficult to explain from the chaotic-atmosphere paradigm,
could be an indication that the weather-climate system is close to a ``critical'' state.
Indeed, Peters and Neelin \cite{Peters_np} 
found evidence of the existence of a sudden but continuous transition
between a non-rainy and a rainy phase as a function of the water-vapor content of the atmosphere, analogous to a thermodynamic second-order phase transition.
In addition, 
the state of the atmosphere would show a tendency to be located at the onset
of this transition, i.e., at the critical point,
which would explain the prevalence of scale invariance in atmospheric processes, 
as scaling is one of the hallmarks of critical phenomena \cite{Stanley_rmp}.
In particular, the scale-invariance found in single-site rain measurements, 
spatio-temporal rain clusters, and tropical cyclones may be a
direct manifestation of Peters and Neelin's findings \cite{Peters_dragon_kings}.

The spontaneous criticality of the atmosphere may be originated by a feedback 
mechanism that 
triggers the existence of %%has 
an attractor at %%
the onset of the transition;
%%as an attractor, 
this phenomenon is referred to as self-organized criticality \cite{Bak_book,Corral_Elsner}.
In simple terms, when the atmosphere is in the subcritical phase
(low water-vapor content and no rain) 
the mechanisms at work increase the water-vapor content, until the critical point is reached and the chance of rain increases;
this hinders the further increase of the water-vapor content.
On the other hand, if the system enters into the supercritical phase (high water-vapor content)
the dynamics is rainy, which decreases the water-vapor content,
until the non-rainy (subcritical) phase is reached.
In this way, the system fluctuates around the critical point.
One key characteristic of criticality is that perturbations evolve 
keeping a delicate balance between 
%attenuation and 
amplification and attenuation
\cite{Zapperi_branching}, 
which has obvious implications for predictability. 
The coexistence and compatibility of this hypothetical criticality of the atmosphere with its chaotic dynamics
remains a fundamental open question.

Nevertheless, 
the concept of scale invariance is problematic from the empirical point of view
\cite{Avnir,Leitao}, 
as it is very difficult to establish its existence rigorously.
For power-law distributions in particular, 
researchers have traditionally used linear regression in double logarithmic scale to fit either
the probability density or the complementary cumulative distribution function, 
but this procedure is known to lead to important biases \cite{Bauke,White,Clauset}.
In consequence, many claims in the literature about the existence of power-law behavior
or self-similarity can be considered dubious.

%Another important problem arises for 
The problem continues when dealing with 
power-law tails, 
where the power-law distribution holds asymptotically, 
or, in practice, above an unknown lower cut-off.
In that case, it has been usual to establish the value of the cut-off ``by the naked eye,''
which is of course arbitrary and irreplicable.
Clauset et al. \cite{Clauset} proposed an ad-hoc method to find the value of the lower cut-off, but other researchers have found this method to perform bad
for synthetically generated power-law tails \cite{Corral_nuclear,Voitalov_krioukov}.
Another common problem is that, for the most extreme events, 
power-law distributions can be perturbed by finite-size effects \cite{Corral_garcia_moloney_font}
and the tail of the distribution transforms into an exponential-like decay
(Clauset et al.'s method \cite{Clauset} and other fitting methods are unable to deal with power laws tapered in this way).
Further, power laws can be easily confused with lognormal distributions
\cite{Malevergne_Sornette_umpu,Corral_Arcaute,Corral_epidemics_pre},
and there are no appropriate tools for model comparison when different distributions hold
above different lower cut-offs 
(as each distribution fits a different subset of the data \cite{Serra_Corral_Zipf}).

The problem of power-law-tail fitting is far from being rigorously solved,
but here we will use the approach explained in Refs. \cite{Corral_Deluca,Corral_Gonzalez}, 
which has given reasonably good results in many different applications.
We obviate the problem of correlations between the observations
\cite{Gerlach_Altmann_prl,Moore_Altmann_PRX}, 
looking for the distribution that best explains
the data values assuming independence
(note that independence is the maximum-entropy outcome when no constrain 
is available for data dependence \cite{Broderick}).
When one has an ergodic system spanning a time window much longer than 
the correlation time, one recovers the probability density of the variable under
study. 
If not, what one obtains in this way is not 
an estimation of 
the (marginal) probability density of the underlying population but 
the probability density conditioned to the observed history and constrains.

A different way to approach the problem of extreme events is by means of extreme-value theory, and, in concrete, by the peaks-over-threshold framework \cite{Coles}.
A limit theorem 
(analogous in some sense to the generalized central limit theorem \cite{Mantegna_Stanley}) 
ensures that, for sufficiently large thresholds (or cut-offs) $u$,
and under statistical independence, the exceedance $x-u$ of a random variable $x$
with respect the threshold follows the so-called generalized Pareto distribution (note that the threshold $u$ introduced here is different to the threshold in the signal mentioned above to the define events). 
%described by the variable $x$). 
The generalized Pareto distribution extends the ``classic'' Pareto distribution to zero and negative values of the so-called extreme-value index $\xi$
(which constitutes the shape parameter of the distribution).
In this way, one gets a mathematical justification to use a particular distribution to fit extremes
(but only for ``extremely'' large thresholds, in theory).
An important problem then in extreme value theory is to find proper methods to establish the value of the required threshold in order that the exceedances follow the generalized Pareto distribution.

%In this work we will use... ??? \cite{Morina_R}
%the mean-excess plot,
%the CV plot (based of the coefficient of variation, CV),
%the Hill plot...?
%Y el log-CV plot??? \cite{Malevergne_Sornette_umpu,Corral_epidemics_pre}
%EXPLICAR BIEN!!

% \cite{Morina_R}
%log-CV plot??? \cite{Malevergne_Sornette_umpu,Corral_epidemics_pre}

%MENCIONAR LOS METODOS PARA ENCONTRAR EL UMBRAL!!!!

Interestingly, if instead of calculating the exceedances with respect to the threshold
one calculates the relative change, i.e., 
one rescales the random variable by the value of the threshold, as $x/u$, 
the resulting limit distribution is not the generalized Pareto but (under certain circumstances) the power law.
%, see the Appendix.
%
%
In fact, the generalized Pareto distribution with positive extreme-value index
and the power-law distribution are very much related: asymptotically, 
the former is characterized by a power-law tail, and both belong to the so-called Fr\'echet maximum domain of attraction in extreme value theory.
Further, when a power-law-distributed variable $x$ is shifted by a constant 
(such as $x-u$) it becomes Pareto distributed.
Informally, we can refer to power-law tails and Pareto tails as ``fat tails.''
The present paper discusses in detail the important relations between both distributions.
Note, however, that extreme-value theory does not provide (in contrast to self-organized criticality)
an explanation for the origin of power-law tails,
as the Pareto distribution arises for distributions that have a power-law tail before
calculating exceedances
(i.e., those distributions belonging to the Fr\'echet maximum domain of attraction \cite{Coles}).
Alternative explanations for the origin of power laws in different systems 
are explained in Refs. \cite{Sornette_critical_book,Mitz,Newman_05,Penland}.

Although, empirically, the fitting of the generalized Pareto distribution and the fitting of the power-law distribution face similar problems 
(the unambiguous and automatic finding of a reasonable cut-off or threshold $u$),
the methodologies employed in each case have been different, 
due probably to the fact that they are used by different communities of researchers.
As already mentioned, automatic ad-hoc algorithms are currently used for power laws by the
statistical physicists and complex-systems researchers
\cite{Clauset,Corral_Deluca,Corral_Gonzalez,Voitalov_krioukov}, 
whereas visual methods 
(diverse plots such as the Hill plot, the mean-excess plot or the CV plot) 
have been traditionally used for the generalized Pareto in extreme-event statistics
\cite{Coles,Morina_R}.

% \cite{Barabasi_criticism}

We investigate in this paper if the important atmospheric phenomenon known as the Madden-Julian (MJ) oscillation (MJO) reflects in some degree the scale-invariance present in many other aspects of weather and climate.
This is important as criticality may underlie or influence
the complexity of the dynamics of MJO propagation.
Moreover, the statistics of MJ events may be affected by different atmospheric conditions and processes.
Our approach is similar to that used in Ref. \cite{Corral_hurricanes}
to study tropical cyclones, but note that the MJO has the additional complication
of being characterized by a phase (in addition to an intensity).
%, which we also investigate. ???
%From the methodological point of view,
We pay special attention to the probabilistic description of the phenomenon
as well as to the fitting procedures.
We align with the ``ad-hoc recipes'' used in complex systems to fit power-law distributions \cite{Corral_Deluca,Corral_Gonzalez}, which have the advantage of being automatized, not visual.
%with the approaches used in extreme-value theory to fit the generalized Pareto distribution.

%As a by-product, we obtain a comparison of the performances of the different approaches to fat-tail fitting.
%NOO!!!???

%PENDIENTE!!!2
In the next section we briefly 
%introduce the MJO, 
explain the index used to quantify the occurrence of the MJO
as well as 
our definition of MJ events.
In Sec. 3 we introduce the two main probability distributions to fit MJ event sizes and durations,
the power-law and the Pareto distributions;
we also mention the important relations between both distributions
(explained in more detail in an Appendix).
In Sec. 4 we present our statistical results, 
showing the convenience to use the double power-law distribution (an extension of the simple power law) 
to characterize the size and durations of the MJ events.
The double power law also makes it clear that there is a sudden decrease
in the survival of the MJO after about 27 days.
Conclusions are presented in the last section.

\section{The MJO, the RMM Index, and definition of MJ events}

\subsection{The MJO through the RMM Index}

The Madden-Julian Oscillation \cite{lin2004stratiform} 
constitutes the principal mode of variability in the tropical weather on sub-seasonal time scales
(this goes from two weeks to approximately three months \cite{Vitart2017TheST}), 
and has a strong influence on the precipitation in the tropics, 
but it also affects higher latitudes through teleconnection patterns \cite{ZhangLing17}.  
%The whole structure
%MJO
The MJO is an atmospheric structure that 
has a tendency to
move eastward 
%from the Indian Ocean 
with an average speed of about $5$ m s$^{-1}$ ($\simeq 4^\circ$ day$^{-1}$) \cite{lin2004stratiform}.
It is characterized by a region of strong convection, with precipitation and upward motion, and ahead (to the east) and behind (to the west) there are regions of suppressed convection with dry conditions.
% Este párrafo yo lo pondría abajo de la descripción del diagrama fase.
Previous studies have found substantial variability in the occurrence of the MJO.
From the physical point of view, 
the basic mechanisms behind MJO are not well understood \cite{Randall_book};
there are several models and hypothesis 
\cite{Zhang_theories_mjo}
but consensus is not reached  \cite{Hottovy_CAFE}.

%PENDIENTE!!!1 !!!!!!!!!!!!!!!!!!!!!!!CITAS!!!!!!!!!!!!!!!!!!!!!
%CITAR RANDALL 2012???

%MEDIR LA DISTRIB DE VELOCIDADES!!!! A LA COLET???

Unambiguous definition of the MJO is elusive \cite{Hand_mjo}, but
to practically monitor it, Wheeler and Hendon \cite{wheeler2004all}
developed a Real-time Multivariate MJO (RMM) index 
which consists of the first and second 
principal components (RMM1 and RMM2) 
obtained from the empirical orthogonal functions (EOF) 
that combine latitudinal averages of outgoing longwave radiation (OLR), and zonal winds at lower (850 hPa) and higher (200 hPa) atmospheric levels. 
The EOFs are calculated for daily fields on a latitudinal band of $\pm 15$ degrees around the Equator. 
Notice that, in contrast to other ``activation'' phenomena, 
the RMM index characterizes the MJO through a bivariate signal,
which introduces an extra degree of complication in comparison 
with univariate signals \cite{Corral_hurricanes,Baro_Corral}.
%and using a 20 to 100 day bandpass filter
%(tengo que revisar bien aquí si es 20 to 100 o 20 to 200 day??).
%MONICA: BIEN, PERO SI NO ESTA EN EL PAPER DE WH HABRA QUE PONER LA CITA CORRESPONDIENTE!!
%%The OLR has a spatial resolution of $2.5 \times 2.5$ degree latitude-longitude global grid. 
%%MONICA: CUAL ES LA RESOLUCION DEL VIENTO??
%%

We download the daily values of the RMM index 
from the Australian Government Bureau of Meteorology, 
from January 1979 to  December 2021
\cite{mjo_index_data}
(records prior to 1979 are found to be incomplete).
Other MJO indices have been developed, but the one by Wheeler and Hendon
(the RMM index \cite{wheeler2004all}) is the most widely used.

%SE PUEDEN APROVECHAR LOS DATOS ANTERIORES A 1979??

\begin{figure}[ht]
\includegraphics[width=.8\columnwidth]{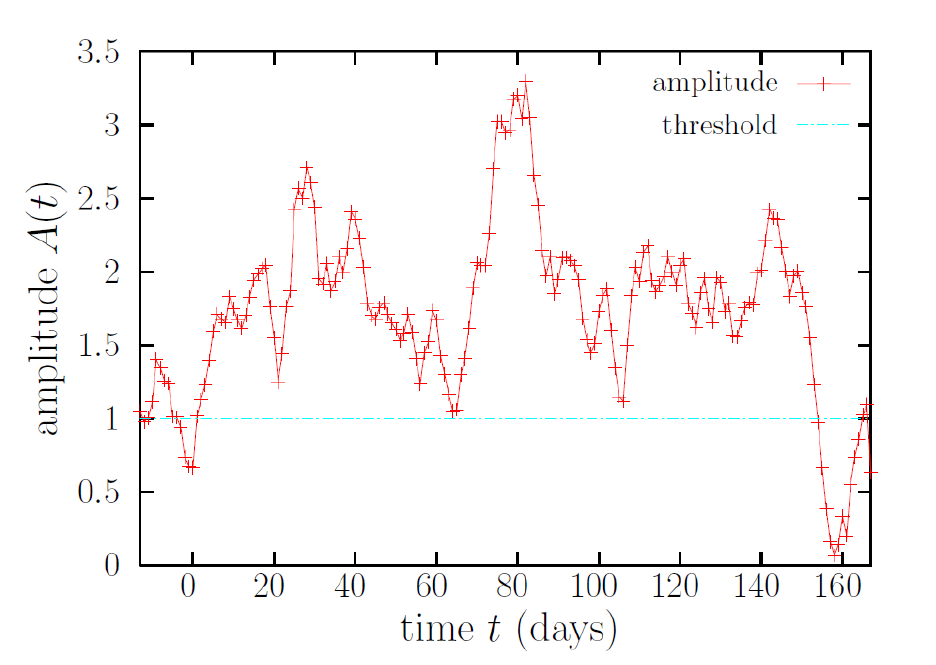}\\
\includegraphics[width=.8\columnwidth]{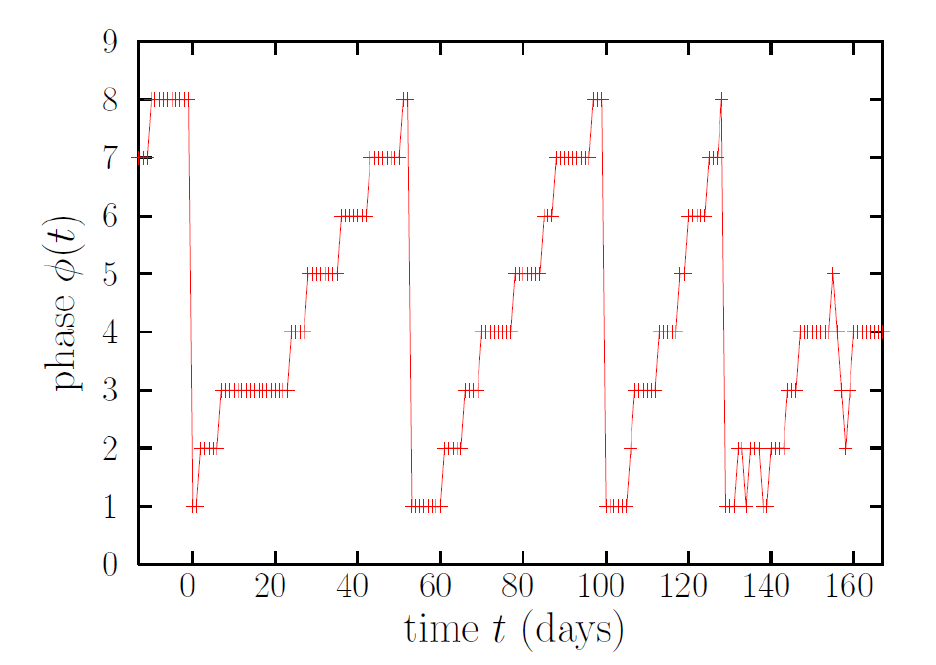}\\
\caption{
(a) Time window displaying the amplitude $A(t)$ of the MJO as given by the RMM index,
including the largest event on record, with $d=153$ days, $s=291$, and 
$n=27$ (i.e., 3.4 cycles).
$t=1$ signals the starting time of this event, 
corresponding to Dec 20, 1989.
(b) Same for the (discretized) phase $\phi(t)$.
}
\label{Fig_amplitude}
\end{figure}

The progression of the MJO can be visualized in a 2-dimensional phase diagram 
with RMM1 in the horizontal axis and RMM2 in the vertical axis.
%see Fig. \ref{pdiagram}(a).
%MONICA: EL NOMBRE CORRECTO ES PHASE DIAGRAM??
Equivalently, the RMM index can be represented in polar coordinates by its amplitude and phase
(note that we use the term phase in the Introduction with a different meaning).
The amplitude $A$, or intensity, is just the modulus of the vector defined by RMM1 and RMM2
in Cartesian coordinates,
i.e.,  
$$
A=\sqrt{RMM1^2 + RMM2^2}.
$$
When the amplitude is above a specific threshold $A_c$, taken equal to one
(i.e., when the vector is outside the unit circle),
% in the RMM1-RMM2 plane) 
the MJO is active
and the vector usually moves counterclockwise.
%describing rough circles reflecting the eastward movement of the MJO. 
If the index is below the threshold then the MJO activity is considered weak 
or suppressed, and the path of the vector is more erratic. 
An example of the MJ amplitude is shown in Fig. \ref{Fig_amplitude}(a).

The MJO phase $\phi$ as given by the RMM index 
can be obtained in several steps, starting with 
$$
\phi=\mbox{arctan2}{(RMM2,RMM1)},
$$ 
where $\mbox{arctan2}$ is the 2-argument arctangent
(yielding continuous values between $-\pi$ and $\pi$).
The resulting phase is shifted and rescaled as
$$
\phi \rightarrow \frac{8(\phi+\pi)}{2\pi},
$$
and subsequently discretized in eight values from 1 to 8,
as 
\begin{equation}
\phi \rightarrow 
%\lceil\phi\rceil=
\lfloor\phi\rfloor +1,
\label{discretization}
\end{equation}
%excluding the trivial case of an integer $\phi$, 
with $\lfloor\phi\rfloor$ the integer part of $\phi$.
In this way, the negative quadrant ($RMM1, RMM2 <0$)
corresponds to phases 1 and 2, and so on, counterclockwise.

The resulting eight discretized phases 
%coincide with 
are associated to
the geographical location of the MJO along the tropical region. 
In particular, 
phases 2 and 3 correspond to the Indian Ocean, 
phases 4 and 5 to the Maritime Continent, 
phases 6 and 7 to Western Pacific, and 
phases 8 and 1 to the West Hemisphere and Africa.
Thus, the usual counterclockwise movement of the vector reflects
the eastward movement of the MJO when it is active.

%When the amplitude of the index is above a specific theshold equal to one
%(i.e., when the vector is outside the unit circle in the RMM1-RMM2 plane) the MJO is strong 
%
%and the vector usually moves counterclockwise describing rough circles reflecting the eastward movement of the MJO. 
%
%If the index is below the threshold then the MJO activity is considered weak and the path of the vector is more random.  
%This threshold is typically considered as 1. 

\subsection{Madden-Julian events, event duration, and event size}

%In our definition, 
In the most simple definition,
based on the RMM index,
an MJ event 
%takes place when the index amplitude is above a threshold. 
starts when the amplitude $A$ of the index crosses the threshold $A_c$ from below
(from $A < A_c$ to $A\ge A_c$, signaling initiation), 
and ends when the amplitude crosses the threshold from above
(from $A\ge A_c$ to $A< A_c$, extinction or completion), 
with the threshold fixed to $A_c=1$
(this prescription is standard although somewhat arbitrary).
Thus, the event consists of all the consecutive days in which the amplitude
is above (or at) threshold, signalling a continuously active MJO.
The example in Fig. \ref{Fig_amplitude} displays the largest MJ event on record
(in terms of largest size and duration).
If an event starts, supposedly, on the first day on record, 
or it has not ended on the last day, 
the event has to be removed from the analyses, 
as it is likely that these are incomplete events.
%Although it is standard to consider the fixed threshold equal one, 
%this prescription is somewhat arbitrary.

It is worth mentioning that the MJ events we define are different from other prescriptions in the literature
(the reason is that one may use the same term for quite different things); 
for instance, Samarasinghe et al. \cite{Barnes_causal_MJO}
consider daily ``events'' 
(each day the amplitude is above the threshold is a different event),
whereas Matthews \cite{Matthews08} identifies ``events'' with complete cycles.
In particular, our definition of MJ events is purely statistical, 
and we do not introduce, a-priori, any arbitrary prescription about which should be, for instance, the minimum duration of an event to be considered a genuine MJ event.
Instead, we rely on the statistical analysis to clarify these issues.
Alternatively,
we could have also used the terms MJ ``instances'' or MJ ``excursions'',
but, for the sake of simplicity, we stick to MJ events.

%We look for the days with amplitude above 1 and we assign a value of 1 and we call them active days. 
%The days with amplitude less than 1 are assigned a 0 and we called them inactive days as shown in figure \ref{active}.

The number of consecutive days with amplitude $A\ge A_c$  
gives the duration $d$ of the MJ event,
computed as $d= t_\text{f}-t_\text{i}+1$ (in days), 
where $t_\text{i}$ is the starting time of the event (first day above threshold),
and $t_\text{f}+1$ is the ending time ($t_\text{f}$ is the last day above threshold).
The size $s$ of the event is the sum of the amplitudes along the duration of the event, i.e.,  
$$
s=\sum_{t=t_\text{i}}^{t_\text{f}} A(t),
$$
with $t$ denoting time (in days).
This is essentially the same definition used for the energy of hurricanes 
\cite{Corral_hurricanes} and the size of rain events \cite{Peters_pre}. 
In the rest of the paper we will study the statistics of both $d$ and $s$,
but with a preference for $s$, for reasons that will become clear later.
Other observables characterizing the events will be defined and used for very concrete purposes below.

%COMPARAR DURACION DEL EVENTO: EN DIAS O EN NUMERO DE FASES DISTINTAS!!!

%DECIR QUE LOS EVENTOS FINALES E INICIALES SE TIRAN???

\section{Power-law and Pareto fittings and their relations}

%\subsection{Power-law and Pareto fittings}

%Our main concern is in the distribution of MJ-event sizes $s$, 
%described by their probability density $f(s)$.

As mentioned in the Introduction, 
we will base our study in two main probability distributions
to fit the empirical values of the MJ event sizes.
Redefining $s$ or $d$ (or any other variable) as $x$, 
the probability density of the power-law (pl) distribution is
\begin{equation}
f_\text{pl}(x) = \frac\alpha a \left(\frac a x\right)^{\alpha+1}
\mbox{ for } x \ge a,
\label{pl}
\end{equation}
and $0$ otherwise, with $\alpha>0$ and $a>0$,
being $\alpha+1$ the exponent of the density
(and $\alpha$ the exponent of the complementary cumulative distribution function)
and $a$ the lower cut-off.
Note that the fit will be performed in such a way that $a$ can be larger
than 
%the lowest values 
some of the values
of $x$, thus, those values will be discarded.

Given a threshold $u$, we define the exceedances $y$ as $y=x-u$, when $x\ge u$
(discarding the rest of values).
Extreme-value theory \cite{Coles} guarantees that the exceedances follow the generalized Pareto (gp) distribution when $u\rightarrow\infty$ and the values of $x$ are independent. The corresponding probability density is
\begin{equation}
f_\text{gp}(y)=\frac 1 {\sigma(1+\xi y /\sigma)^{1+\frac 1 \xi}}
\mbox{ for } y \ge 0,
\label{gp}
\end{equation}
and $0$ otherwise, considering $\xi\ge 0$ and $\sigma>0$,
being $1+\xi^{-1}$ the exponent (of the density)
and $\sigma$ a scale parameter.
$\xi$ is referred to as the extreme-value index
(not be confused with the extremal index). 
% https://projecteuclid.org/journals/bernoulli/volume-11/issue-6/Estimation-of-the-extreme-value-index-and-generalized-quantile-plots/10.3150/bj/1137421635.pdf
% Beirlant et al. Bernoulli 2005
In fact, $\xi$ can be smaller than zero 
(and in that case $0 \le y \le \sigma/|\xi|$),
but the case of interest for us is $\xi\ge 0$,
where the particular case $\xi=0$ corresponds to an exponential distribution,
whereas $\xi > 0$ corresponds to the standard (non-generalized) Pareto distribution.

Of special relevance, but not particularly for extreme events, is the truncated power-law (tpl) distribution, which may arise when, in a power-law distribution, the most extreme values
deviate from the power law.
Then, in order to avoid additional parameters to model this deviation, these
most extreme events, above some upper cut-off $b$ 
(that truncates the distribution from above), are eliminated.
The distribution is
defined by the probability density
\begin{equation}
f_\text{tpl}(x)=\frac {\alpha a^\alpha}{1-(a/b)^\alpha} \left(\frac 1 x\right)^{\alpha+1}
\mbox{ for } a \le x < b,
\label{tpl}
\end{equation}
and zero otherwise, 
with $-\infty <\alpha<\infty$ but $\alpha\ne 0$ and $0 < a < b$.
The parameters $\alpha$ and $a$ play the same role as in the (non-truncated) power-law distribution, 
but note that they can take very different values
(as in our case every distribution will fit a different range of the data);
thus, we will distinguish $a_\text{tpl}$ and $\alpha_\text{tpl}$
from $a_\text{pl}$ and $\alpha_\text{pl}$.
The upper truncation parameter $b$ gives name to the distribution and
yields the non-truncated case in the limit $b\rightarrow \infty$ if $\alpha>0$.
In the next section we will see the convenience, from an empirical point of view, 
of introducing, in addition, the double power-law distribution.

%\subsection{Relation between the power law and the Pareto distributions}

The connections between the power law and the Pareto distribution ($\xi >0$)
are explained in the Appendix I. A practical summary follows:

\begin{itemize}
%1
\item Pareto is asymptotically a power law, i.e.,
$f_\text{gp}(y) \rightarrow f_\text{pl}(y)$
when $y \rightarrow \infty$,
with $\alpha=1/\xi$ and $a=\sigma/\xi$.

%2
\item
If $x$ is power law,
the exceedances $y=x-u$ are Pareto, with
$\xi=1/\alpha$ and $\sigma=u/\alpha$, if $u\ge a$.

%3
\item If $y$ is
Pareto, the shifted variable $x=y+u$ is power law
if $u=\sigma/\xi$, with $\alpha=1/\xi$ and $a=u$. 

\item
If $y$ is Pareto (for $y\ge 0$), the shifted variable $x=y+u$ is ``shifted Pareto'' for any value of $u$
(and for $x\ge u$), with the same values of $\xi$ and $\sigma$.

%4
\item
The $k-$th order moments, 
$\langle x^k\rangle$ and $\langle y^k\rangle$ 
(corresponding to power law and Pareto, and also for shifted Pareto), 
become infinite (diverge)
for $k\ge \alpha=\xi^{-1}$ (assuming $\xi > 0$).

%5
\item Pareto is an attractor
for a broad class of distributions
when $y=x-u$ and $x>u$,
with $u\rightarrow \infty$ and independent values of $x$.

\item
The power law is an attractor when $z=x/u$ and $x>u$, 
with $u\rightarrow \infty$ 
and independent values of $x$.

\end{itemize}

%Thus, we conclude that there is a certain equivalence between fitting a power law to some data and fitting a Pareto distribution to its exceedances, 
%in concrete, a power law always implies a Pareto distribution for the exceedances, whereas the reciprocal is true if the shift $u$ and the cut-off $a$ of the power law are precisely selected.
%Additionally, a power law can be theoretically justified in the same way as a Pareto distribution,
%just considering $x/u$ instead of the exceedances $x-u$.
%%In the next section we will see the convenience, from an empirical point of view, 
%%of introducing the double power-law distribution
%%FALTA UNA FORMULA ILUSTRATIVA AQUI???
%

\section{Results}

%\subsection{Elementary statistics}

\begin{figure}[ht]
\includegraphics[width=.8\columnwidth]{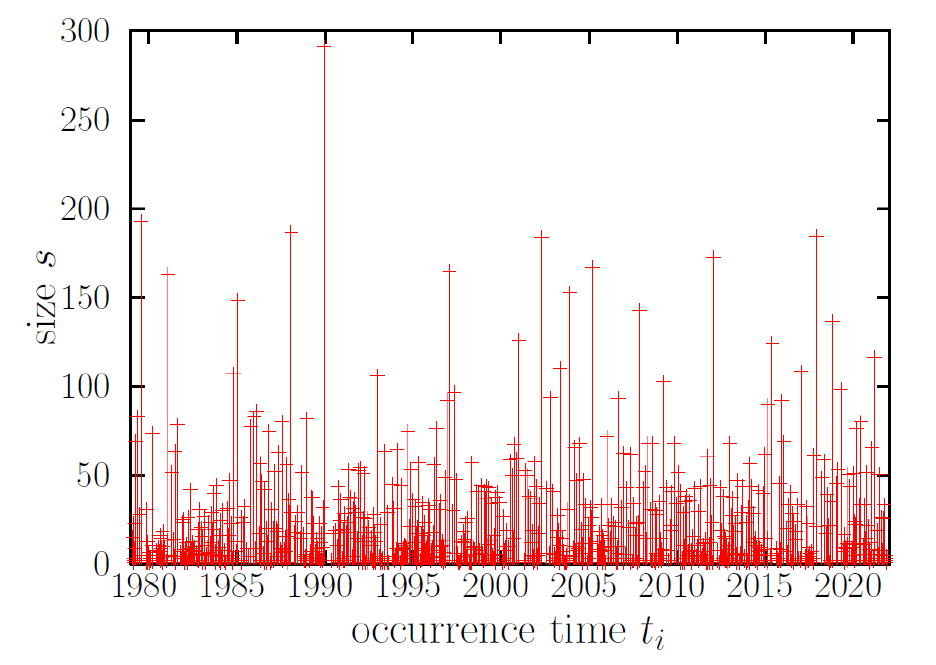}
\caption{
%(a) Time series of number of Madden-Julian events per year.
%(b) 
Point process corresponding to the size $s$ of MJ events at their occurrence times,
given by the starting time $t_\text{i}$.
%INCLUIR BARRAS DE ERROR!!
%PONER TAMBIEN LO MISMO PARA LOS EVENTOS ENCIMA DE 39???
%POCOS DATOS?? PONER 20???
%(b) Number of MJ events per month, normalized by total number of events
%(representing therefore a probability distribution),
%compared with number of active days per month (normalized by total number also).
}
\label{Fig_event_numbers}
\end{figure}

With the prescriptions explained in the previous section
we obtain a total of
734 MJ events during the period 1979-2021 (43 years), 
which yields a rate of
17.1 MJ events per year.
%In Fig. \ref{Fig_event_numbers}(a) we observe the annual variability of the number of MJ events.
For comparison, the average number of active days per year turns out to be
226
(corresponding to 62~\% of the days and a mean duration $\langle d \rangle\simeq 13.2$ days).
In Fig. \ref{Fig_event_numbers} we display the values of the size in time
(as a marked point process).

%DURACION MEDIA 13.2 DIAS???
%
%In order to assess the seasonality of MJ-event occurrence, 
%we compute  
%the number of MJ events per month 
%(of different years, where each event is assigned to its starting month),
%displayed in Fig. \ref{Fig_event_numbers}(b),
%where it is compared with the number of active days per month
%(both normalized by their total numbers).
%%The mean of both quantities is 61 (events per month)
%%and 810 (active days per month).
%A negative correlation of the two different counts is apparent, 
%which is confirmed by a negative value of the Pearson coefficient 
%%between both variables 
%($-0.6$).
%In short, the first half of the year is characterized by more active days and less events,
%and the opposite for the second half of the year, 
%which %can be interpreted as if 
%implies that
%in the first half of the year MJ events have longer duration than in the second half
%(later we will verify that this is indeed the case). 
%%VERIFICAR!!!! COMPARAR ESTADISTICA jfmM VERSUS JAS!!!!

\subsection{Probability distributions and scatter plots}

\begin{figure}[ht]
\includegraphics[width=.8\columnwidth]{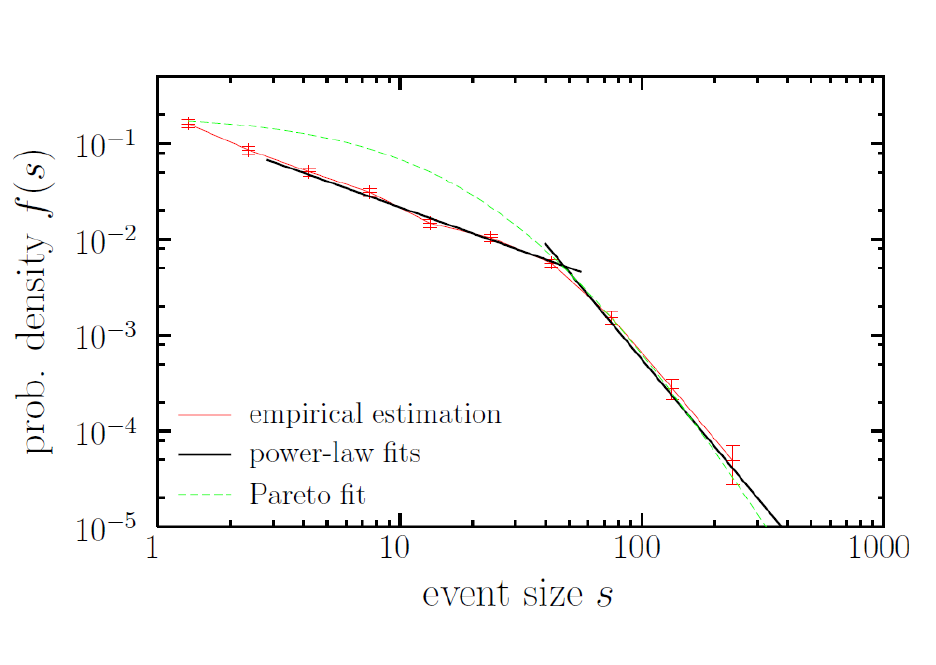}\\
\includegraphics[width=.8\columnwidth]{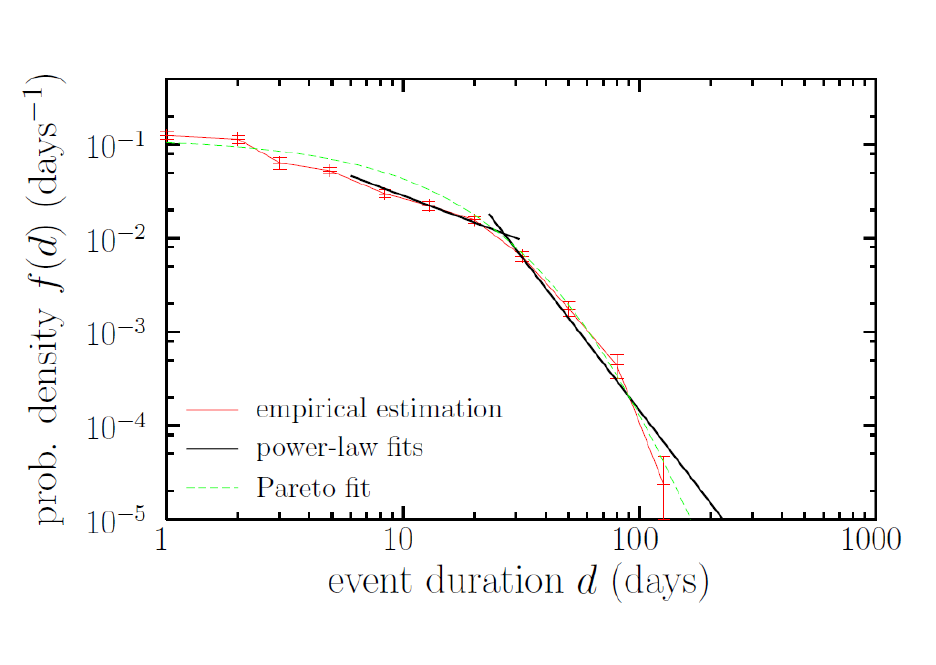}
\caption{
(a) Empirical estimation of the probability density $f(s)$ of the size of MJ events
(in log-log scale).
Power-law and Pareto fits are also shown,
with exponents 
$\alpha_\text{tpl}=-0.10$,
$\alpha_\text{pl}=2.0$,
and
$\alpha_\text{gp}=3.2$
(b) Same for the probability density $f(d)$ of the durations.
Power-law and Pareto fits are given by
$\alpha_\text{tdpl}=-0.05$, 
$\alpha_\text{dpl}=2.3$,
and
$\alpha_\text{gp}=7.3$
(the power-law fits for $d$ are discrete but represented by lines).
For both observables 
Pareto fits are valid for $s,d\ge a$; 
nevertheless, the region below $a$ is shown for illustration.
}
\label{sizes}
\end{figure}

The estimation of the empirical probability density of the sizes of the MJ events, $f(s)$, 
using the full record,
appears in Fig. \ref{sizes}(a) (in double logarithmic scale),
showing its broadness, ranging from $s=1$ to almost 300.
The counterpart for event durations, $f(d)$, shown in Fig. \ref{sizes}(b),
turns out to be qualitatively similar, 
ranging from 1 day to more than 150 days.
%Figure \ref{rescaling} rescales both distributions in order to compare them, 
%confirming the qualitative similarity in shape.
%For the sake of completeness,
%both distributions are shown again in Fig. XXX in terms of their survivor function 
%(or cumulative complementary distribution function).
%
%RESCALADO NAIVE O A LA ROSSO?? rosso!!!!
%
Notice that the distribution of $f(d)$ shows no special behavior around its smallest values, 
which means that durations $d=1$, 2, or 3 are not ``pathological'', 
and constitute part of the same phenomenon given by the longer durations.
The only remarkable change is around $d\simeq 27$ days 
(see Fig. \ref{sizes}(b), and below for the quantification),
where a jump in the slope (in log-log) is apparent
(this corresponds to an event size around 47, and will be more precisely quantified below).

%PARECE QUE SERIAN 47 EN VEZ DE 50!!! CAMBIAR!!!!

\begin{figure}[ht]
\includegraphics[width=1.\columnwidth]{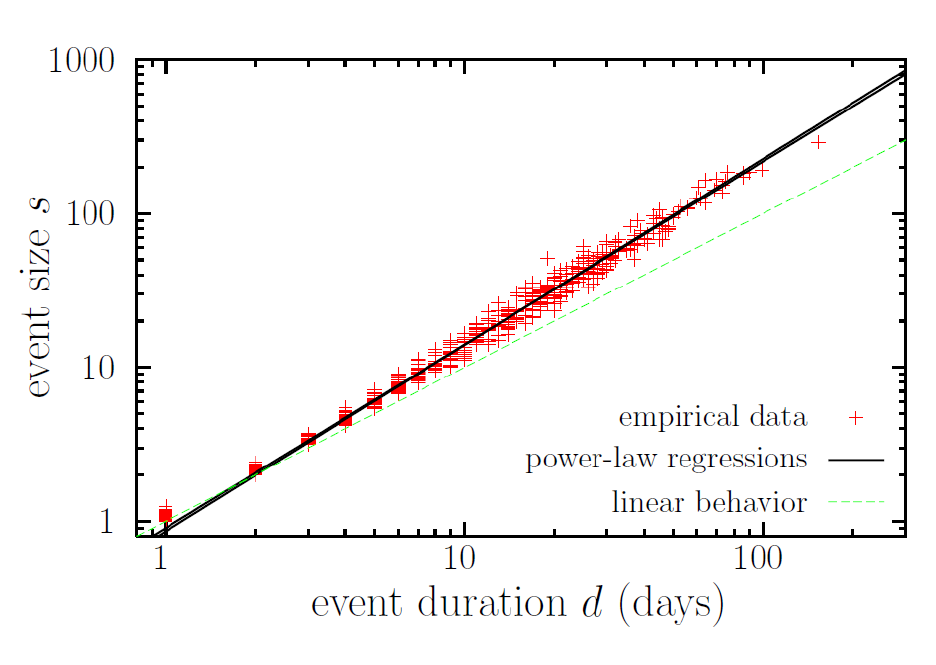}
\caption{
Scatter plot (in log-log) showing the relation and high correlation between sizes and durations of MJ events.
Note that the relation is non-linear, in concrete power law 
with an exponent $\gamma \simeq 1.2$ 
(the two regression lines are nearly the same).
Linear relation also shown, for comparison.
}
\label{scatter}
\end{figure}

The correlation between the size and the duration of the events is displayed in the scatter plot of Fig. \ref{scatter}.
We fit a straight line to $\ln s$ versus $\ln d$ 
%in the range $d>3$,
(excluding the events with $d \le 3$, to reduce discreteness effects),
leading to the power-law relation $s \propto d^\gamma$
with $\gamma=1.191 \pm 0.007$ and a Pearson correlation coefficient 
$\rho=0.992$
(the opposite regression, that of $\ln d$ versus $\ln s$ leads to $\gamma=1.211 \pm 0.007$).
We refer to this linear correlation between the logarithms as 
power-law correlation.
Observe that no change in slope (in log-log) is observable, 
despite the marginal distributions $f(d)$ and $f(s)$ show it.
%a change in slope.
Thus, the relation between $s$ and $d$ seems to be more ``fundamental'' 
than the individual marginal distributions, 
as the $s-d$ relation is maintained for the full range of values of $s$ and $d$.

We can also compute the maximum amplitude within a MJ event, $A_\text{max}$, 
as well as the mean amplitude $A_\text{mean}$, which is given by $A_\text{mean}=s/d$. 
It turns out that
%(where $s$ is size and $d$ duration).
these two measures of event amplitude 
are highly correlated between them, 
with a linear regression of the logarithms leading to the regression curve
$A_\text{max}\propto  A_\text{mean}^{1.55 \pm 0.01}$ 
%and $A_\text{max}\propto  A_\text{mean}^{1.63 \pm 0.01}$,
and a Pearson coefficient $\rho=0.977$
(exponent 1.63 for the opposite regression).
%(as before, the second power-law curve corresponds to the opposite regression;
%we follow always this convention).
However, in contrast to $s$ and $d$,
these two variables are not broadly distributed, 
as they are restricted to a range 
from 1 (the threshold value) to no more than 5 
(the empirical upper bound to $A_\text{max}$);
in consequence,
we do not study the corresponding probability distributions.
Nevertheless, the two amplitudes are also highly correlated with size and duration.
In the same way, we find $A_\text{max}\propto s^{0.280\pm 0.007}$ 
%and $A_\text{max}\propto s^{0.358\pm 0.007}$ (opposite regression) 
for $s\ge 5$ with $\rho=0.885$
(the exponent for the opposite regression is 0.358),
as well as
$A_\text{mean}\propto s^{0.175\pm 0.005}$ 
%and $A_\text{mean}\propto s^{0.252\pm 0.005}$ 
for $s\ge 5$ with $\rho=0.83$
(exponent 0.252 for the opposite regression).

\subsection{Power-law and Pareto fittings of the event sizes}

We proceed by fitting a power-law tail, Eq. (\ref{pl}), to the distribution of MJ-event sizes.
%using, in principle, the complete data set (all seasons). 
%to have better statistics, 
%as Fig. \ref{seasons}
%makes it clear that the differences in slope (in log-log) between the different seasons
%should be small.
As it was obvious from the visual inspection of Fig. \ref{sizes}(a), 
a power law cannot fit all events
(we do not observe a complete straight line in a log-log plot), 
and we need to find a value of the lower cut-off $a$ for which all events above $a$ are well fitted, disregarding the rest.
This is a way in which one fits a power-law tail, 
and this is the idea of Clauset et al.'s method \cite{Clauset},
although, instead, as mentioned in the Introduction,
we will use the alternative method exposed in Refs. \cite{Corral_Deluca,Corral_Gonzalez}, 
which is similar in spirit but has been found to yield more consistent results in controlled tests \cite{Corral_nuclear,Voitalov_krioukov}.
%MOVER LA FRASE DE ARRIBA??

Requiring, being very strict, 
a $p-$value larger than 0.20 to accept (i.e., not reject)
a power-law fit over a specified range, 
when applied to the distribution of sizes $f(s)$ this method leads to
a good fit for
$a_\text{pl}=40$ (comprising 131 events),
with
$1+\alpha_\text{pl}=3.0\pm 0.2$
and a $p-$value $=0.85$.
% 3.03 y 39.8
%
As we know,
this power-law fit is equivalent to a (generalized) Pareto distribution for the exceedances $y=x-a$ with 
$\xi_\text{pl}=1/\alpha_\text{pl}=0.5$ 
and $\sigma_\text{pl}=a_\text{pl}/\alpha_\text{pl}=20$.
However, 
if instead of a power law we fit in the same range a Pareto distribution to the exceedances,
Eq. (\ref{gp}), we get
$\xi_\text{gp}=0.3\pm 0.1$ (corresponding to $1+\alpha_\text{gp}=4.2$)
and $\sigma_\text{gp}=23\pm 3.5$.
This fit is performed by using the function {\it gpd} from the R-package {\it evir}.
The performance of both fits can be seen at Fig. \ref{sizes}(a).

%DECIR COMO SE AJUSTA LA PARETO!!!! 

%DECIR QUIEN GANA CON AKAIKE!!! 

The reason of the difference (not discrepancy), in particular in the value of the exponents
$\alpha_\text{pl}$ and $\alpha_\text{gp}$, 
is that, once the value of $a$ is selected, 
the power law only has one parameter to fit, 
whereas the Pareto distribution has two,
allowing for a ``better fit'' (at the cost of being less parsimonious).
With this case we illustrate the possibility of very different outcomes for the power-law
and the Pareto fits, despite their asymptotic equivalence:
Due to the fact that the fitted Pareto distribution has not reached its asymptotic behavior in the tail of the empirical distribution, its exponent becomes substantially 
larger than that of the power law. 
In this regard, 
the value of the exponent brought by the Pareto distribution
can be difficult to interpret, 
as, in some sense, the exponent is not directly visible in the data
due to the resulting value of the scaling parameter $\sigma$
(the power-law tail contained in the Pareto distribution only would be visible for $s > 300$).

%Estaria bien DERIVAR ESTE 300 DE LOS PARAMETROS!!!!!!!

%DECIR QUE TAMBIEN SE HA FITEADO LA LN, PERO ES KK!! OJO, NO tan KK!!!! da mejor rango que la PL y la GPD. Fitea desde infinito hasta 10.

The relative performance of both fits
can be quantified using the Akaike information criterion 
(AIC$=2 P-2\ln L$, with $P$ the number of parameters and $L$ the likelihood).
For the power law, AIC$_\text{pl}=1170.54$,
whereas for the Pareto (or shifted Pareto) AIC$_\text{gp}=$AIC$_\text{sp}=1170.80$;
thus, the highest likelihood provided by the Pareto is not enough to 
beat the parsimony of the power-law distribution and we conclude that the power law
is a better fit. 
If instead of the AIC we use the Bayesian information criterion, 
the advantage provided by the power law becomes higher.
% del gnuplot
%pareto(x)=1/sigma/(1+xi*(x-u)/sigma)**(1+1/xi)
%sigma = 23.02
%xi = 0.313
%u = 39.8

%We apply now several methods used in extreme-value theory to find the threshold $u$ to fit the generalized Pareto distribution.
%
%MOVER LO DE ABAJO??
%These methods are
%the Hill plot, 
%the mean-excess plot,
%and the CV plot.
%In fact the Hill plot takes advantage of the fact that the tail of a Pareto distribution is asymptotically a power law and approximates the tail index
%of the Pareto by the inverse of the maximum-likelihood value of the exponent 
%$\alpha$.
%
%Note that both the mean excess plot and the CV plot are based on the calculation of moments, which do no exist for some values of $\xi^{-1}$.
%In concrete, the mean excess plot is not valid for $\xi^{-1} \le 1$
%and the CV plot for $\xi^{-1} \le 2$,
%so, extreme caution is required here.
%
%Y EL LOG-CV PLOT???

%COMPARAR TODOS LOS AJUSTES EN REPRESENTACION PARETO... 
%Y TODOS LOS POWER LAW??? (QUIERE DECIR PARE EXCEEDENCES???)

%PONER TAMBIEN LOS FITS PARA LA ACUMULADA!!!

In order to try to characterize the full distribution of sizes and not only the tail,
we attempt the fit of a truncated power law, Eq. (\ref{tpl}), to some range of the size data.
The fitting method is essentially the same as for the untruncated power law, 
see Refs. \cite{Corral_Deluca,Corral_Gonzalez}
(in contrast, Clauset et al.'s method \cite{Clauset} is unable to fit truncated power laws).
We obtain a good fit with an exponent $1+\alpha_\text{tpl}=0.90\pm 0.05$ in a range
from $a_\text{tpl}=3$ to $b_\text{tpl}=56$ (comprising 488 MJ events)
with a $p-$value $=0.30$.
Note that the resulting value of the exponent $\alpha_\text{tpl}$
is close to the value obtained for tropical cyclones \cite{Corral_hurricanes},
another phenomenon governed by convection;
so, one may wonder if what we are seeing in MJO is the other side of the same coin.

\subsection{The double power-law distribution}

The truncated-power-law fit together with the previous fits of the tail (power law or Pareto)
cover all the data, except the smallest values ($s <3$), 
which have to be disregarded.
In particular, a double power law
(combining the truncated power law for the body of the distribution with 
the untruncated power law for the tail \cite{Serra_Corral_Zipf})
seems a particularly satisfying solution as (given $a_\text{tpl}$) 
it only involves three parameters
($\alpha_\text{tpl}$, $\alpha_\text{pl}$, and the crossover point between the two regimes, 
given by a value of $s$ around $\theta=47$, as seen in Fig. \ref{sizes}(a)).
The impossibility to fit very small sizes ($s<3$)
may be due, in addition to the very large number of small events
(requiring much more precision in the fits),
to the fact that $s$ has a strange character as a random variable, 
as it originates from the integration of a continuous one (the amplitude)
along discrete time, and this has a clear signature for small events
(for larger events, $s$ becomes continuous, in practice).

The probability density of the double power law (2pl) can be considered a
mixture of the (untruncated) power law and the truncated power law, 
Eqs. (\ref{pl}) and (\ref{tpl}), respectively, 
weighted by a parameter $q$
(and with both distributions defined over different supports), i.e.,
\begin{equation}
f_\text{2pl}(x)=(1-q) \frac{\alpha_1}{\theta} \frac 1 {(\theta/a)^{\alpha_1}-1} 
\left(\frac \theta x \right)^{\alpha_1+1} \mbox{ for } a\le x \le \theta,
\label{doublepl1}
\end{equation}
\begin{equation}
f_\text{2pl}(x)=     q \frac{\alpha_2}{\theta}                                     
\left(\frac \theta x \right)^{\alpha_2+1} \mbox{ for } x\ge  \theta,
%\end{alignat*}
\label{doublepl2}
\end{equation}
and zero otherwise.
The exponents fulfill $-\infty<\alpha_1<\infty$ but with $\alpha_1\ne 0$
and $\alpha_2 > 0$. 
The scale parameter $\theta$ fulfills $\theta \ge a$
(and marks the sudden change of slope in log-log)
and the lower cut-off $a$ fulfills 
$a\ge 0$ if $\alpha_1< 0$ and
$a> 0$ if $\alpha_1> 0$.
Power laws have no characteristic scales, 
but a double power law has one, given by $\theta$;
so, in some sense, $\theta$ allows to introduce a non-arbitrary separation
between ordinary events and extreme events.
Notice that $q$ is not a free parameter but ensures continuity between the two power-law regimes by requiring 
$q=\alpha_1/[\alpha_2 (\theta/a)^{\alpha_1}-(\alpha_2-\alpha_1)]$.
We identify 
$\alpha_1=\alpha_\text{tpl}$, 
$\alpha_2=\alpha_\text{pl}$
and $a=a_\text{2pl}=a_\text{tpl}\ne a_\text{pl}$.
In the ideal case, $b \simeq \theta \simeq a_\text{pl}$
but in practice we calculate $\theta$ from the intersection of the two power-law regimes,
as seen in Fig. \ref{sizes},
see Ref. \cite{Serra_Corral_Zipf} also.

Due to the good performance of the double power-law distribution in the fitting of the event-size data, and the power-law correlation between event size and the rest of the variables studied, 
in the remaining of the article we will pay special attention to the double power-law distribution
(as we explain below, 
a power-law regime for one variable 
together with a power-law correlation with a second variable 
leads to another power-law regime for the second variable).
Summarizing, the results of the double power-law fit 
for the distribution of the sizes of MJ events
are
$a_\text{2pl}=3$
(encompassing 558 events),
$1+\alpha_{1}=0.93 \pm 0.05$,
$1+\alpha_{2}=3.0 \pm 0.2$,
and $\theta=47$.

%MAS??? cuantos eventos incluye????

\subsection{Fitting of the event durations}

We proceed by applying the same approach to the distribution of event durations, $f(d)$.
However, the results of the power-law fitting are bad in this case, in particular for the 
truncated power-law regime. 
The reason is that the duration $d$ is a discrete random variable 
(measured in number of days)
and the fitted power-law distributions are continuous
(as mentioned, the size reflects in part this discretization, 
but the size is not a pure discrete variable, as the duration).
Therefore, we simply redefine the power-law distributions to deal with discrete random variables, which essentially only changes the normalization constant in Eq. (\ref{pl}),
see Ref. \cite{Corral_Deluca_arxiv,Corral_Cancho}
(the normalization constant is fundamental in maximum-likelihood estimation).

The resulting probability mass function (the equivalent to the probability density)
of the truncated discrete power law (tdpl) turns out to be
$$
f_\text{tdpl}(x)=\left[\frac 1 {\zeta(\alpha+1,a)-\zeta(\alpha+1,b+1)}\right]\, \frac 1 {x^{\alpha+1}}
\mbox{ for } x=a,a+1,\dots,b
$$
and zero otherwise, with $\zeta(\nu,c)=\sum_{x=c}^\infty 1/x^\nu$ the Hurwitz zeta function, 
$0 < a \le b$
and
$-\infty < \alpha <\infty$ for $b$ finite (tdpl)
but 
$\alpha > 0$ if $b\rightarrow \infty$,
leading to $\zeta(\alpha+1,b+1) \rightarrow 0$
and defining the (untruncated) discrete power-law (dpl) distribution.

%PONER FORMULA POWER LAW DISCRETA???

% ojo al histograma, tiene que ser con 5 cortes por decada!!
The results obtained when fitting the (untruncated) discrete power law 
to the event durations
are 
$a_\text{dpl}=23$ days
(comprising 141 events), % p 0.42
$1+\alpha_\text{dpl}=3.3 \pm 0.2$,
and $p=0.42$. 
On the other hand, a Pareto fit leads to 
$\xi_\text{gp}=0.14\pm 0.10$
(corresponding to a very large $1+\alpha_\text{gp}=8.3$)
and
$\sigma_\text{gp}=14 \pm 2$.
%
% result par.ests
%        xi       beta 
% 0.1367416 13.9158248 
% result par.ses
%       xi      beta 
% 0.1014116 1.8611351 
The Akaike information criterion leads to AIC$_\text{pl}=1036.8$ for the power law
and AIC$_\text{gp}=1039.8$, so the power law outperforms the Pareto
(approximating the power law to the continuous case)
and is therefore preferred.
% o sea, cogiendo los parametros de la PL discreta y metiendolos en la formula de la continua.
The truncated discrete power law (for fiting beyond the tail) leads to
$a_\text{dtpl}=6$, 
$b_\text{dtpl}=31$ days (with 370 events),
$1+\alpha_{dtpl}=0.95 \pm 0.12$,
and $p=0.71$. % p 0.80 % promediando cuatro veces sale 0.71.
As the two power-law regimes overlap, 
the double power law constitutes a good description, 
for which
%When taking the double-power-law description 
we obtain a change of slope at
$\theta=27$ days, see Fig. \ref{sizes}(b).
Thus, the discrete double power-law is able to fit the event durations for $d\ge 6$ days with exponents 0.95 and 3.3.
The fact that the Pareto distribution yields such an extremely large value of the exponent (8.3) may be an indication of the lack of stability of the fit when the asymptotic regime is not reached and of the superiority of using the simpler power law for the tail.

%Y AKAIKE!! %% Y FALTAN P VALUES ARRIBA!

\subsection{Relations and correlations between different observables}

As the sizes and durations are correlated, 
the exponents of both power-law distributions are not independent. 
Remember that we have shown (in a certain sense, see Fig. \ref{scatter}) that
$s\propto d^\gamma$, with $\gamma\simeq 1.2$.
There is a well known theoretical relation \cite{footnote_mjo2,Corral_thesis}
between the exponent $\gamma$ and the power-law exponents, 
given by 
\begin{equation}
\gamma=\frac{\alpha^\text{(d)}}{\alpha^\text{(s)}}=\frac{\xi^\text{(s)}}{\xi^\text{(d)}}
\label{exponent_rel}
\end{equation}
(where we have introduced the superscripts to distinguish between $s$ and $d$, of course).
Certainly, 
when we consider the untruncated power-law tails,
the relation is fulfilled by the empirical values we have obtained
($\alpha_\text{pl}^\text{(s)}\simeq 2.0$ and 
$\alpha_\text{pl}^\text{(d)}\simeq 2.3$), within statistical uncertainty.
This is a further reason to prefer the power-law fit in front of the Pareto fit, 
as for Pareto we would obtain 
${\alpha_\text{gp}^\text{(d)}}/{\alpha_\text{gp}^\text{(s)}}
=7.3/4.2=1.7$, far from the empirical value of $\gamma$.
For the truncated power laws describing the bulk of the distributions,
the values of $\alpha_\text{tpl}$ are very close to zero, and the associated uncertainties turn out to be larger than the values of $\alpha_\text{tpl}$, 
making impossible a proper validation of the relation between the power-law exponents and $\gamma$.

%PARA PARETO TAMBIEN FUNCIONARIA LA RELACION???
%PARECE QUE ES UN CHURRO!!!
%es culpa del exponente de pareto para las duraciones!!!

In addition, 
the relation between the power-law exponents and $\gamma$
given by Eq. (\ref{exponent_rel})
can explain why the amplitudes $A_\text{mean}$ and $A_\text{max}$ are 
not broadly distributed. 
Take for instance $A_\text{mean}$, for which we have established $s\propto A_\text{mean}^\gamma$ with $\gamma\simeq 4$ (see above). 
As $\alpha_\text{pl}^\text{(s)}\simeq 2$; 
this would lead to an hypothetical power-law exponent $\alpha_\text{pl}^\text{(Am)}\simeq 8$ 
for $A_\text{mean}$,
which implies a very fast decay, very difficult to detect empirically and to distinguish from an exponential decay, with the limited number of data that we have.
This constitutes indirect evidence that both $A_\text{mean}$ and $A_\text{max}$ 
could be power-law distributed, but with very high values of the exponents.

\subsection{Increased probability of extinction of MJ events and conditional distributions}

As we have seen,
the different fits of $f(s)$ and $f(d)$ confirm a clear change of behavior for intermediate values of $s$ and $d$, 
which means that MJ events beyond a ``barrier'' of about $\theta=27$ days (47 units in size) have more difficulties to survive.
Where does this barrier come from?
As the phase conveys spatial information, 
it seems interesting to separate
the size and duration distributions into different starting and ending phases
of the MJ events.

\begin{figure}[ht]
\includegraphics[width=.8\columnwidth]{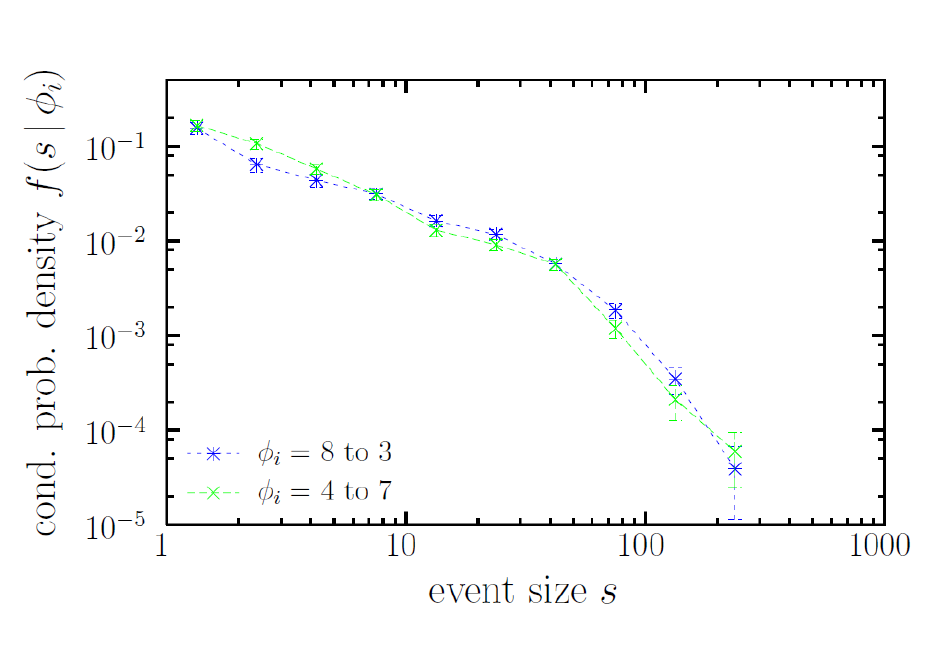}\\
\includegraphics[width=.8\columnwidth]{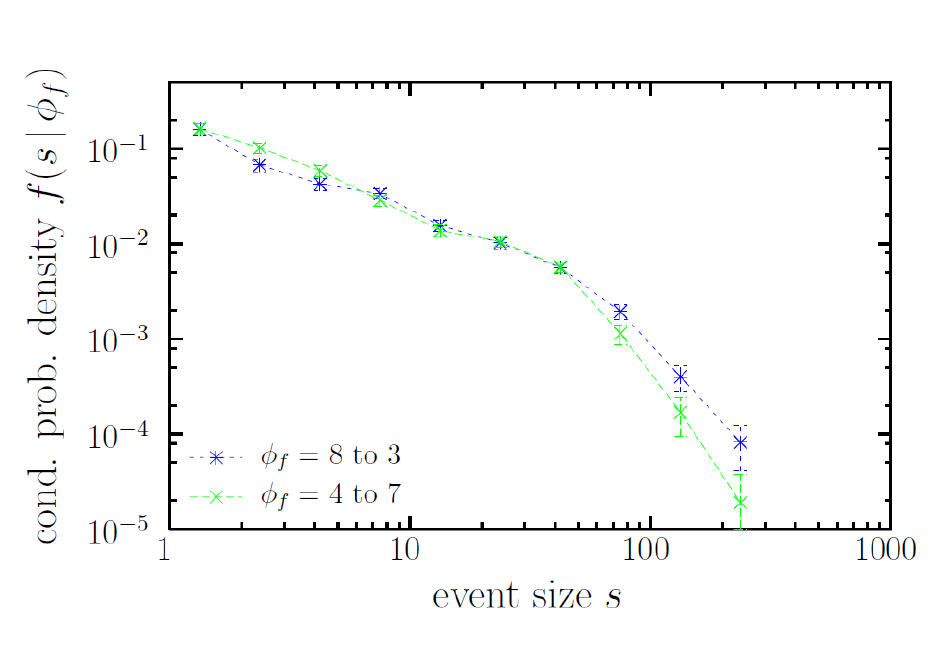}
\caption{
(a) Empirical probability densities of event sizes restricted to different values of their initial phases.
The two subsets have nearly half of the events each.
No clear influence of the initial phase on the change of slope is observed.
(b) Empirical probability densities of sizes restricted to their final phases.
The number of events is 
353 for $\phi_\text{f}$ from 8 to 3
and 
381 for $\phi_\text{f}$ from 4 to 7.
There seems to be an influence in the tail, 
%but the uncertainty there is large. 
but we show in the main text that it is not significant.
}
\label{Dsizes_phasein}
\end{figure}

Thus, we consider  
the probability density of event size conditioned to a given set of values of the starting phase, 
$f(s\,|\, \phi_\text{i})$,
or conditioned to (a given set of values of) the ending phase,
$f(s\,|\, \phi_\text{f})$.
The starting phase $\phi_\text{i}$ of an event is its phase at the starting time $t=t_\text{i}$,
and the ending phase $\phi_\text{f}$ is the phase at $t=t_\text{f}$. 
We find that $f(s\,|\, \phi_\text{i})$ shows practically no dependence on the initial phase,
see Fig. \ref{Dsizes_phasein}(a)
(where we compare the size distribution for events starting in phases 8 to 3 with 
that for starting phases from 4 to 7, i.e., 
events starting from the Western Hemisphere to the Indian Ocean with events staring
in the Maritime Continent or the Pacific;
we group the phases because for individual values of the phases the statistics is too low).
In fact, the slight difference between the two conditional distributions
can be explained by the statistical fluctuations arising from the low number of events at the tails
of the distributions.
Changing sizes for durations the conclusion is the same.

%UN POCO LIO ARRIBA, s o d??

On the other hand, $f(s\,|\, \phi_\text{f})$ seems to show a somewhat larger effect of the phase 
(ending phase in this case), in particular for the most extreme events, 
%as those events ending with $\phi_\text{f}=8$ to $3$ have longer durations 
%than extreme events ending with phases in the range from 4 to 7, 
see Fig. \ref{Dsizes_phasein}(b),
where
there are relatively more extreme events ending in phases 8 to 3 than in phases 4 to 7,
which could mean that 
the former (those ending from the Western Hemisphere to the Indian Ocean)
have an increased probability of extinction in comparison with 
the latter (which end at the Maritime Continent or at the Pacific Ocean);
nevertheless, this happens with no apparent change in the value of $\theta$
(remember: the characteristic scale for $s$ or $d$).
As an illustration,
out of the 15 most extreme events with $\phi_\text{i}= 8$ to 3, 
ten end in the same range of phases (after completing one cycle, or more)
and 5 end in the range from 4 to 7.
On the contrary, 
out of the 11 most extreme events with $\phi_\text{i}= 4$ to 7,
only 3 end in the same phase and 8 end with $\phi_\text{f}= 8$ to 3.
In the next subsection we will show that this effect seems to be in fact an artifact, 
and the best explanation is that there is no significant influence of the ending phase
on the sizes and durations of the events, 
i.e., sizes and durations can be considered independent of ending phases
(in the same way they are more clearly seen as independent of the starting phases).

\subsection{Total phase advance of MJ events}

%At this point 
For the purpose of clarifying the previous issue,
we introduce a variable that counts 
%the different number of phases $n$ 
the total phase advance $n$
of a MJ-event, defined as 
$$
n=\sum_{t=t_\text{i}+1}^{t_\text{f}} \Delta \phi^\text{c}(t),
%+ 1, 
$$
where $\Delta \phi^\text{c}(t) =\phi^\text{c}(t)-\phi^\text{c}(t-1)$ 
accounts for the daily changes in the continuous phase $\phi^\text{c}$ 
(the continuous phase is the value of the phase previous to discretization, 
taking values from 0 to $8^-$,
i.e., $\phi=\lfloor \phi^\text{c} \rfloor +1$, see Eq. (\ref{discretization})).
Notice that a change of $\phi$ 
from 8 to 1 has to be counted as  $\Delta \phi=1$
and therefore the resulting $\Delta \phi^\text{c}$ has to be increased in eight
(and the opposite for a change from 1 to 8).
For that reason, in general, $n \ne \phi^\text{c}(t_\text{f})-\phi^\text{c}(t_\text{i})$.
In fact, $n$ turns out to be the number of cycles in an event,
but multiplied by 8.
Notice also that the phase advance can be negative, and a few MJ events 
%defined in this way 
are characterized by negative values of the total phase advance
(the event with the smallest phase change yields $n=-1.25$;
in contrast, the largest value in data is $n=27$, which corresponds to 3.4 cycles, see Fig. \ref{Fig_amplitude}).
%This event observable 

\begin{figure}[ht]
\includegraphics[width=1.\columnwidth]{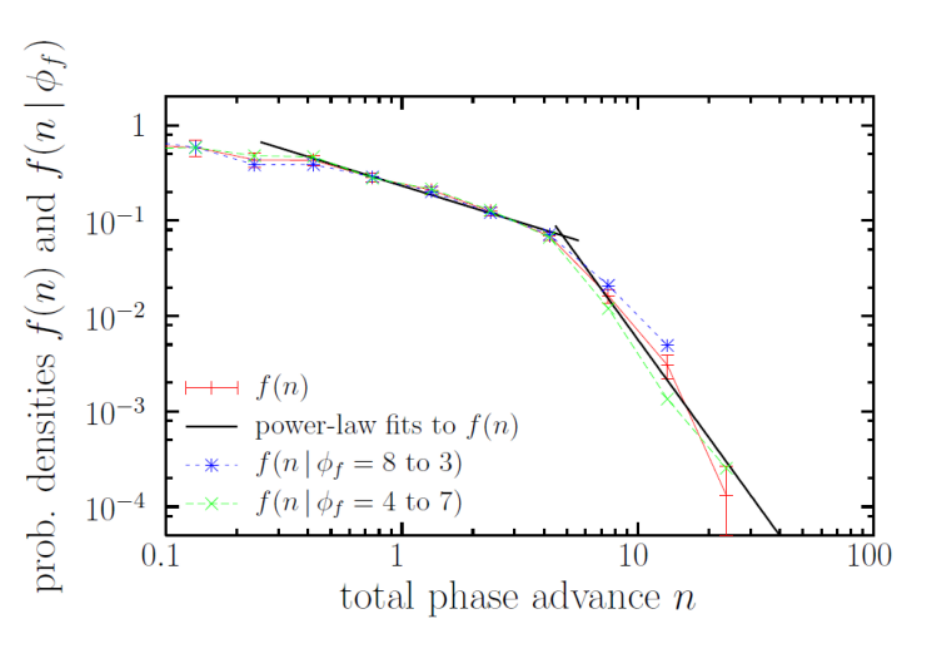}
\caption{
Empirical 
probability density $f(n)$ of the total phase advance of each MJ event 
(where the phase is not discrete but continuous).
Events with $n \le 0.1 $ are excluded.
Power-law fits are also included.
A sudden change of slope around $n=\theta=5$ is observed.
Conditional distributions $f(n\,|\,\phi_\text{f})$ are also included.
%Distribucion exponencial??
}
\label{nphases}
\end{figure}

The total phase advance
$n$ turns out to be power-law correlated with the size $s$
(and therefore with the rest of observables studied in this work).
Indeed, we find 
$s\propto n^{0.83\pm 0.02}$ 
%$s\propto n^{1.13\pm 0.02}$ 
for $n\ge 0.1$ 
(1.13 for the opposite regression)
with $\rho=0.856$.
Figure \ref{nphases} shows that the probability density $f(n)$ of $n$ is
(as expected from the power-law correlation with $s$)
broadly distributed.

%HACER LA REGRESION INVERSA Y COMPROBAR QUE EL ERROR ES EL MISMO!!!!
% no puede ser el mismo, zoquete!!

The fit of $f(n)$  
yields a power-law tail given by
$a_\text{pl}=4.5$
(comprising 90 events), % p 0.33
$1+\alpha_\text{pl}=3.4 \pm 0.25$,
and $p=0.32$. 
The fit of a truncated power law leads to
$a_\text{tpl}=0.25$, 
$b_\text{tpl}=5.6$ (comprising 421 events),
$1+\alpha_{tpl}=0.77 \pm 0.055$ % p 0.22
and $p=0.22$.
The two power laws cross at a value of $n=\theta=5$,
which means that after a total phase advance of about 5
%(where 8 represents a complete cycle)
(a bit more than half a cycle)
the MJ event is more likely to decay,
which would indicate a sort of ``exhaustion'' of the MJO,
in the same way as reflected by the size and duration of the events.

%AQUI ARRIBA ESTA LA CLAVE!!!

Separation of $f(n)$ into different starting phases
leads to distributions very similar to $f(n)$ (not shown);
%is shown in Fig. \ref{Dn_cond},
%yielding a plot similar to Fig. \ref{Dsizes_phasein},
some small difference is observed for different starting phases
but the difference does not seem significant in comparison with the uncertainty in the estimation of the distributions.
The probability densities conditioned to different ending phases 
are included in Fig. \ref{nphases},
leading to results analogous to those for 
$f(s | \phi_\text{f})$,
in which it is difficult to discern if there is an effect of the ending phase.

Some simulations are helpful at this point.
We want to generate independent values of 
the staring phase $\phi_\text{i}$
and of the total phase advance $n$
(due to the independence inferred from $f(n\,|\,\phi_\text{i})$),
and for this purpose
we resample their empirical distributions. 
%of staring phases $\phi_\text{i}$
%and of total phase advances $n$.
Assuming 
%as indicated by Fig. \ref{Dsizes_phasein}(a), 
that $\phi_\text{i}$ and $n$
are independent we resample both distributions independently, 
i.e., we take a value uniformly from the list of empirical values of  $\phi_\text{i}$,
and do the same (independently) from the values of $n$, and calculate the
corresponding
final phase $\phi_\text{f}$ from both variables.
%as $\phi^\text{c}_\text{f}=(\phi_\text{i}+n)mod$.
In this way we are able to compute the distribution of $n$ conditioned to 
the final phase, as well as the distributions of $s$ and $d$
(in this case, the resampled values of $s$ and $d$ are not taken independently from $n$,
as they are correlated, so, corresponding values need to be taken).

We find that when we repeat this procedure many times 
(very large number of resampling, e.g., 100,000)
the resulting distributions conditioned to final phase turn out to be indistinguishable between them
(do not depend on  $\phi_\text{f}$).
This demonstrates, numerically, that independence of $n$, $s$, and $d$ on 
the initial phase implies independence or nearly independence on the ending phase 
(for a distribution of initial phases given by the empirical distribution).
In contrast, when the number of resamplings is small, 
in concrete when it is the same as the observed number of events $N$,
the tail of the distributions may show apparent differences between them, 
which therefore we conclude are just statistical fluctuations.
%This confirms that the ending phase does not have a significant influence on the distributions
%of $s$, $d$, and $n$.

%\begin{figure}[ht]
%%\includegraphics[width=1.\columnwidth]{figs/fig_Dn_cond_phasefin.pdf}
%\caption{
%$f(n)$ conditioned to final phase!!!
%FALTARIA LA INICIAL??
%INCLUDE SIMULATIONS??
%}
%\label{Dn_cond}
%\end{figure}

%$$
%h_\text{2pl}(x)=
%\frac{(1-q)\alpha_1 \theta^{\alpha_1}}
%{\left[q\left( \theta/a\right)^{\alpha_1}-1\right] x ^{\alpha_1+1}
%+(1-q) \theta^{\alpha_1} x} 
%\mbox{ for } a\le x \le \theta,
%$$
\subsection{Peak of the hazard rate}

The increased probability of extinction of the MJO when the characteristic scale $\theta$ is reached
can be understood better using the hazard rate function $h(x)$ \cite{Kalbfleisch2}.
This is defined as the probability density, but conditioned to the fact that the random variable has indeed reached a value $x$
(so, $h(x)$ is not normalized by the total number of events
but it is ``normalized'' by the number of events that reach the value $x$, which depends, of course, on $x$), 
and is given by $h(x)=f(x)/S(x)$, with $S(x)$ the complementary cumulative distribution function.
For the double power-law distribution, Eqs. (\ref{doublepl1}) and (\ref{doublepl2}), we obtain  
$$
h_\text{2pl}(x)=
\frac{(1-q)|\alpha_1| \theta^{\alpha_1}}
{\left[1-q\left( \theta/a\right)^{\alpha_1}\right] x ^{\alpha_1+1}
-(1-q) \theta^{\alpha_1} x} 
\mbox{ for } a\le x \le \theta,
$$
$$
h_\text{2pl}(x)=\frac{\alpha_2}{x} \mbox{ for } x\ge  \theta,
$$
taking the expression for $S(x)$ from Ref. \cite{Corral_Gonzalez}
and using that $\alpha_1< 0$.
The denominator (in the first formula) has a maximum at
$$
\sqrt[\alpha_1]{
%\left\{
\frac{(1-q)\theta^{\alpha_1}}
{[1-q(\theta/a)^{\alpha_1}](\alpha_1+1)}
}
%\right\}^{1/\alpha_1} 
%\simeq 26.9,
$$
corresponding to a size equal to 27.2 when we introduce the parameters describing the size of the events
and to a duration of 14.9 days substituting the corresponding parameters.
This means that the hazard rate, as given by
$h_\text{2pl}(x)$ reaches a minimum at that value and
therefore 
$h_\text{2pl}(x)$ has positive derivative at $s\rightarrow \theta^- $, 
but negative derivative at $s\rightarrow \theta^+ $;
then the hazard rate has a maximum at $s=\theta$;
this constitutes a (local) maximum of the ``hazard'' that the MJO becomes extinct, 
in other words, the extinction rate is peaked at $x=\theta$
(with $x$ representing size or duration).

\section{Conclusions}

We have defined Madden-Julian events from the RMM index, 
and have studied the statistics of several of their observables.
Sizes and durations of MJ events 
present very high variability, being broadly distributed, 
and are quantitatively well described by double power-law distributions
(except for the smallest and shortest events).
The power-law tail contained in the double power-law distribution outperforms 
a (generalized) Pareto fit of the tail, both for sizes and durations.

We have found some inconsistencies when fitting the Pareto distributions:
although extreme-value theory teaches us that the generalized Pareto distribution holds
above (infinitely) large thresholds (assuming independence),
the resulting Pareto fits do not show their asymptotic behavior
(the tail of the Pareto distribution is not seen in the range covered by the empirical data).
Further, the Pareto exponents obtained for both sizes and durations are incompatible with the clear
scaling relation between both variables.
This supports the preference for the double power law.

The double power-law fits make it clear that, 
for durations less than about 27 days,
a MJ event propagates without a characteristic scale for extinction,
as given by the first power-law regime (parameterized by $\alpha_1$) 
of the size and the duration distributions;
however, for the events that reach 27 days
(which, obviously, constitutes a characteristic scale, breaking the scale invariance)
the MJ propagation becomes more difficult in comparison, and extinction more likely.
We have investigated, from a statistical point of view, the origin of this sharp increase in the probability of extinction, 
conditioning the distributions (of sizes and durations) to different starting and ending phases
(taking advantage that the phase of the MJO is related to its position along the Equator).
The resulting conditional distributions do not show any significant influence neither of the starting phase
nor of the final phase (this latter case is less clear and some resampling of the empirical data is necessary to clarify the issue).

Ruling out the influence of the phase implies that the influence of the position can be ruled out as well;
for example, a MJ event starting in the Indian Ocean and another one in the Pacific 
are practically indistinguishable in terms of their duration and size.
This means that we cannot associate the increased extinction of the MJ events to 
the effect of the Maritime-Continent barrier or to any other geographical aspects.
In some sense, all MJ events, independently of their starting position, see the same ``barrier'' after about 27 days, 
and thus, we speculate that this effect, instead of an external influence, is due to an intrinsic cause,
related to the MJO dynamics.
As the MJO is considered to be one of the most important drivers
of sub-seasonal forecasting,
%(the scale going
%from two weeks to three months, roughly \cite{Vitart2017TheST}
the fact that most MJ events do not survive this 27-days barrier
limits the time horizon in which the MJO can bring forecasting skill.

Although tropical cyclones present energy distributions very similar to the size distribution of MJ events
(with values of the exponent $\alpha_\text{tpl}$ very close to one 
and a subsequent deviation from the first power-law regime in both cases),
the extinction of tropical cyclones was considered in Ref. \cite{Corral_hurricanes} to be extrinsic,
caused by the boundary conditions imposed by the finiteness of the different oceanic basins over which tropical cyclones evolve;
this would constitute an important difference between both phenomena.
In a preliminary analysis, we do not see any clear influence of the ENSO and the NAO
in the extinction of the MJO.

%DECIR ALGO DE LA mc BARRIER!!!

%DECIR QUE NO HAY INFLUENCIA DE ENSO O NAO...??? QUE MAS???

\section{Acknowledgements}

We are indebted to Isabel Serra and Pascal Yiou.
This work is part of the Climate Advanced
Forecasting of Sub-Seasonal Extremes (CAFE) project and has
been prepared in the framework of the doctorate in Physics of
the Autonomous University of Barcelona. The authors gratefully
acknowledge funding from the European Union's Horizon
2020 research and innovation programme under the Marie
Sklodowska-Curie Grant Agreement 813844.
The institution of A. C. and M. M. is supported by the 
CERCA Programme of the Generalitat de Catalunya
as well as by the
Spanish State Research Agency
through the Severo Ochoa and Mar\'{\i}a de Maeztu Program for Centers and 
Units of Excellence in R\&D (CEX2020-001084-M).

%support from project
%PGC-FIS2018-099629-B-I00
%from Spanish MICINN
%is acknowledged.

Appendix III provides the codes that are the core to reproduce the main results in this paper.

\section{Appendix I: Relation between the power law and the Pareto distributions}

There are fundamental connections between the power law and the generalized Pareto distribution with $\xi >0$ (i.e., the Pareto distribution):
\begin{itemize}
%1
\item The Pareto distribution yields a power law, asymptotically;
i.e., $f_\text{gp}(y) \rightarrow f_\text{pl}(y)$ for large $y$, 
with $\alpha=1/\xi$ and $a=\sigma/\xi$.
However, note that the power-law character of the Pareto distribution depends on $\xi$ in a very particular way. 
The scale at which the power-law tail is reached is given by $\sigma/\xi$;
therefore, the probability that a Pareto random variable is at its power-law tail
is $C^{-1/\xi}$ (with $C$ some constant larger than 1),
which becomes very small when $\xi$ approaches 0
(and becomes 1 when $\xi\rightarrow \infty$).
In other words, this probability depends on $\xi$ and vanishes for $\xi=0$
\cite{Holme_private}.

%2
\item
Taking the exceedances $y=x-u$ of a power-law distributed variable $x$ yields a  Pareto distribution, with
$\xi=1/\alpha$ and $\sigma=u/\alpha$, if $u\ge a$
(i.e., the exceedances of any power law over any threshold $u\ge a$ yield a Pareto,
as stated by the limit theorem of extreme-value theory,
but the limit $u\rightarrow \infty$ is not required).
Notice that, in contrast with the exponent ($\xi^{-1}$),
the scale parameter $\sigma$ of the resulting Pareto 
barely reflects any property of the original power-law data, 
but strongly depends on the selected threshold.
%3
\item Conversely, 
a Pareto-distributed variable $y$ 
shifted as $x=y+u$ 
yields a power law for $x$ if $u=\sigma/\xi$, with $\alpha=1/\xi$ and $a=u$ 
(i.e., a shifted Pareto yields a power law for a precise value of the shift $u$;
in other words, the power law is a very particular case of a shifted Pareto).
In general, the shifted Pareto (sp) distribution is given by the probability density
$$
f_\text{sp}(x)=
\frac 1 {\sigma}\left(1+\xi \,\frac{x-u} \sigma\right)^{-\left(1+\frac 1 \xi\right)}
\mbox{ for } x \ge u,
$$
and $0$ otherwise, with $-\infty < u < \infty$ (and restricted to $\xi >0$).
\item
Needless to say, if the exceedances $y$ follow a generalized Pareto distribution for $y\ge 0$, the variable $x$ defined as $x=y+u$, for any value of $u$, 
will follow a shifted Pareto distribution for $x\ge u$.

%4
\item The $k-$th order moments 
$\langle x^k\rangle$ and $\langle y^k\rangle$
of both distributions (power law and Pareto), 
do not exist (i.e., become infinite)
for $k\ge \alpha=\xi^{-1}$ (remember that we assume $\xi > 0$).
Of course, this also holds for the shifted Pareto distribution.

%5
\item As we mentioned in the Introduction \cite{Coles}, 
for independent values of $x$
and $x>u$ with $u\rightarrow \infty$, 
the Pareto distribution is an attractor
for a broad class of distributions
under the transformation $y=x-u$.

\item
In a similar way, the power law is an attractor under the transformation $z=x/u$, 
for $x>u$ and $u\rightarrow \infty$ (the distribution of $x$ is not specific,
but, in addition to the power law there are other limiting distributions, 
see the next Appendix).
\end{itemize}

Thus, we conclude that there is a certain equivalence between fitting a power law to some data and fitting a Pareto distribution to its exceedances, 
in concrete, a power law always implies a Pareto distribution for the exceedances, whereas the reciprocal is true if the shift $u$ and the cut-off $a$ of the power law are precisely selected.
Additionally, a power law can be theoretically justified in the same way as a Pareto distribution,
just considering $x/u$ instead of the exceedances $x-u$.
%In the next section we will see the convenience, from an empirical point of view, 
%of introducing the double power-law distribution
%FALTA UNA FORMULA ILUSTRATIVA AQUI???

\section{Appendix II: The power-law as a limit distribution}

%EN ESTA SECCION, EXPLICAR QUE $X/U$ TIENE COMO ATRACTOR LA POWER LAW!!!

Given a random variable $x$ and a threshold value $u$,
let us consider $z=x/u$ for $x>u$;
then, $\ln z = \ln x -\ln u$.
The sometimes-called Pickands–Balkema–De Haan theorem \cite{Coles}
ensures that, for $\ln u \rightarrow \infty$
(i.e., for $u \rightarrow \infty$), a generalized Pareto distribution emerges for $\ln z$
(if some regularity conditions are fulfilled).
Considering the particular case $\xi =0$ 
(note that $\xi=0$ for $\ln z$, not for $z$)
we have that the attractive distribution is the exponential,
defined 
%we have an exponential distribution 
for $\ln z > 0$,
which in terms of $z$ leads to a power law, $f(z)=1/z^{1+1/\sigma}$ for $z>1$, with $\sigma>0$;
this yields another power law in terms of $x$,
$$
f(x) = \frac 1 {u\sigma} \left(\frac u x\right)^{1+1/\sigma}
\mbox{ for } x>u
$$
(and zero otherwise).
Note that $\sigma$, the scale parameter of the exponential distribution for $\ln z$,
becomes the extreme-value index of the power-law distribution for $x>u$
(i.e, the inverse plus one of the exponent).
In some sense, we could define $\xi'=\sigma$
(using standard notation in extreme-value theory).

The case $\xi > 0$ (for $\ln z$) is of less interest for us than the case $\xi=0$,
nevertheless, it yields
$$
f(x)=\frac 1 {\sigma   [1+\sigma^{-1}\xi \ln (x/u)]^{1+1/\xi}\, x}
\mbox{ for } x>u
$$
(and zero otherwise).
This is a regularly varying function with the power-law term $1/x$ 
(with exponent 1, or $\xi'=\infty$) multiplying a slowly varying function.
Notice that this decay is slower than any power-law tail of the type $1/x^{1+1/\xi}$ with $\xi > 0$.
The case $\xi=0$ is included in the case $\xi > 0$ 
taking the limit $\xi\rightarrow 0$.
%This case is of less interest for us than the case $\xi=0$.

The case $\xi<0$ is of no interest for us (as it does not have a power-law tail); 
nevertheless, it is included in the previous formula, 
with the additional constrain $1+\sigma^{-1}\xi \ln (x/u) > 0$. 
%(compact support).
Thus, the previous formula for $f(x)$ holds, 
but, for the sake of clarity, it can also be written as 
$$
f(x)=\frac 1 {\sigma x} \left(1-\frac{|\xi|}{\sigma} \ln \frac x u\right)^{-1 + 1/|\xi|}
\mbox{ for } u<x<u e^{\sigma/|\xi|}
$$
(and zero otherwise).
Moreover, an exponential distribution,
$f(x)=\lambda e^{-\lambda(x-u)}$,
is also possible as a solution, 
but not included in the previous formulas. 
Somehow, the exponential should be in between the cases $\xi <0$ and $\xi=0$.
As far as we know, this solution has no counterpart in the 
Pickands–Balkema–De Haan framework.

In summary, the Pickands–Balkema–De Haan theorem, through a simple transformation, ensures that
the limiting distribution for $x$ when $x>u$ and $u\rightarrow \infty$ is a power law
if the limiting distribution for $\ln x -\ln u$ is an exponential 
(corresponding to a Gumbel maximum domain of attraction for $\ln x$).
If the limiting distribution for $\ln x -\ln u$ is not an exponential,
other attractors arise for $x$. 

\section{Appendix III: Code}

The core of the code that defines events from the daily values of the RMM index is the following
(in Fortran; the threshold $A_c$ is denoted as $u$, the variable $t_\text{f}$ is defined differently as in the main text, but as this is not recorded it has no influence in the results):

\begin{verbatim}
open(10,file='rmm.1979_2021toRealtime.txt',status='unknown')
u=1
Ao=1.e9	  
Nevents=0
do t=1,1e9
read(10,*,end=999) year, month, day, rrm1, rrm2, phase, A 
if ((Ao.lt.u).and.(A.lt.u)) then ! nothing happens
else if ((Ao.lt.u).and.(A.ge.u)) then ! event starts
	   Nevents=Nevents+1
	   ti=t
	   size=A
else if ((Ao.ge.u).and.(A.ge.u)) then ! event continues
	   size=size+A
else if ((Ao.ge.u).and.(A.lt.u)) then ! event ends
	   tf=t 
	   dur=tf-ti
	   if (Nevents.gt.0) print*,ti,dur,size ! to file events_list.dat
endif
Ao=A
enddo
999 close(10)

\end{verbatim}

The probability densities can be estimated from the list of values of the random variable 
(e.g., duration or size) using the following commands (in R):
\begin{verbatim}
xx<-read.table("events_list.dat",header = FALSE); x<-xx$V2
hlog<-hist(log(x),probability = 'T',col='blue')
plot(exp(hlog$mids),hlog$density/exp(hlog$mids),log='xy',type='p')
\end{verbatim}

%\bibliographystyle{unsrt}
%%%\bibliography{biblio}
%%%\bibliography{C:/Users/acorr/Dropbox/p1_lemmas/biblio}
%%%\bibliography{biblio_old,Biblio_mon.bib}
%\bibliography{C:/Users/acorr/Dropbox/p1_lemmas/biblio,Biblio_mon.bib}
%%\bibliography{C:/Users/Store/Dropbox/p1_lemmas/biblio,Biblio_mon.bib}

\begin{thebibliography}{10}

\bibitem{Ben_Zion_review}
Y.~Ben-Zion.
\newblock Collective behavior of earthquakes and faults: continuum-discrete
  transitions, progressive evolutionary changes, and different dynamic regimes.
\newblock {\em Rev. Geophys.}, 46:RG4006, 2008.

\bibitem{Bodenschatz}
E.~Bodenschatz, S.~P. Malinowski, R.~A. Shaw, and F.~Stratmann.
\newblock Can we understand clouds without turbulence?
\newblock {\em Science}, 327(5968):970--971, 2010.

\bibitem{Williams_census}
P.~D. Williams, M.~J. Alexander, E.~A. Barnes, A.~H. Butler, H.~C. Davies,
  Ch.~I. Garfinkel, Y~Kushnir, T.~P. Lane, J.~K. Lundquist, O.~Martius, R.~N.
  Maue, W.~R. Peltier, K.~Sato, A.~A. Scaife, and Ch. Zhang.
\newblock A census of atmospheric variability from seconds to decades.
\newblock {\em Geophys. Res. Lett.}, 44(21):11,201--11,211, 2017.

\bibitem{Ghil_Lucarini}
M.~Ghil and V.~Lucarini.
\newblock The physics of climate variability and climate change.
\newblock {\em Rev. Mod. Phys.}, 92:035002, 2020.

\bibitem{Franzke}
Ch. L.~E. Franzke, S.~Barbosa, R.~Blender, H.-B. Fredriksen, T.~Laepple,
  F.~Lambert, T.~Nilsen, K.~Rypdal, M.~Rypdal, M.~G. Scotto, S.~Vannitsem,
  N.~W. Watkins, L.~Yang, and N.~Yuan.
\newblock The structure of climate variability across scales.
\newblock {\em Rev. Geophys.}, 58(2):e2019RG000657, 2020.

\bibitem{Deluca_Moloney_Corral}
A.~Deluca, N.~R. Moloney, and A.~Corral.
\newblock Data-driven prediction of thresholded time series of rainfall and
  self-organized criticality models.
\newblock {\em Phys. Rev. E}, 91:052808, 2015.

\bibitem{Corral_Elsner}
A.~Corral.
\newblock Tropical cyclones as a critical phenomenon.
\newblock In J.~B. Elsner, R.~E. Hodges, J.~C. Malmstadt, and K.~N. Scheitlin,
  editors, {\em Hurricanes and Climate Change: Volume 2}, pages 81--99.
  Springer, Heidelberg, 2010.

\bibitem{Yang_Franzke}
L.~Yang, C.~L.~E. Franzke, and Z.~Fu.
\newblock Power-law behaviour of hourly precipitation intensity and dry spell
  duration over the {United States}.
\newblock {\em Int. J. Climatol.}, 40(4):2429--2444, 2020.

\bibitem{footnote_mjo1}
When extremes are defined in terms of the values of the variable but not when
  they are defined in terms of percentiles of its probaility distribution.

\bibitem{Peters_pre}
O.~Peters and K.~Christensen.
\newblock Rain: Relaxations in the sky.
\newblock {\em Phys. Rev. E}, 66:036120, 2002.

\bibitem{Corral_Gonzalez}
A.~Corral and A.~Gonz\'alez.
\newblock Power law distributions in geoscience revisited.
\newblock {\em Earth Space Sci.}, 6(5):673--697, 2019.

\bibitem{Peters_Deluca}
O.~Peters, A.~Deluca, A.~Corral, J.~D. Neelin, and C.~E. Holloway.
\newblock Universality of rain event size distributions.
\newblock {\em J. Stat. Mech.}, P11030, 2010.

\bibitem{Deluca_npg}
A.~Deluca and A.~Corral.
\newblock Scale invariant events and dry spells for medium-resolution local
  rain data.
\newblock {\em Nonlinear Proc. Geophys.}, 21:555--567, 2014.

\bibitem{Benzi_21}
R.~Benzi, I.~Castaldi, F.~Toschi, and J.~Trampert.
\newblock Self similar properties of avalanche statistics in a simple turbulent
  model.
\newblock {\em Phil. Trans. R. Soc. A}, 380:20210074, 2021.

\bibitem{Bunde}
A.~Bunde, J.~F. Eichner, J.~W. Kantelhardt, and S.~Havlin.
\newblock Long-term memory: a natural mechanism for the clustering of extreme
  events and anomalous residual times in climate records.
\newblock {\em Phys. Rev. Lett.}, 94:048701, 2005.

\bibitem{Corral_csf}
A.~Corral.
\newblock Scaling in the timing of extreme events.
\newblock {\em Chaos. Solit. Fract.}, 74:99--112, 2015.

\bibitem{Corral_fires}
A.~Corral, L.~Telesca, and R.~Lasaponara.
\newblock Scaling and correlations in the dynamics of forest-fire occurrence.
\newblock {\em Phys. Rev. E}, 77:016101, 2008.

\bibitem{Morina_storms}
D.~Mori{\~n}a, I.~Serra, P.~Puig, and A.~Corral.
\newblock Probability estimation of a {Carrington-like} geomagnetic storm.
\newblock {\em Sci. Rep.}, 9:2393, 2019.

\bibitem{Peters_dragon_kings}
O.~Peters, K.~Christensen, and J.~D. Neelin.
\newblock Rainfall and dragon-kings.
\newblock {\em Eur. Phys. J. Spec. Top.}, 205(1):147--158, 2012.

\bibitem{Traxl}
D.~Traxl, N.~Boers, A.~Rheinwalt, B.~Goswami, and J.~Kurths.
\newblock The size distribution of spatiotemporal extreme rainfall clusters
  around the globe.
\newblock {\em Geophys. Res. Lett.}, 43(18):9939--9947, 2016.

\bibitem{Lovejoy81}
S.~Lovejoy.
\newblock Analysis of rain areas in terms of fractals.
\newblock In {\em Proc. 20th Conference on Radar Meteorology}, pages 476--483.
  American Meteorological Society, 1981.

\bibitem{Lovejoy}
S.~Lovejoy.
\newblock Area-perimeter relation for rain and cloud areas.
\newblock {\em Science}, 216:185--187, 1982.

\bibitem{Emanuel_nature05}
K.~Emanuel.
\newblock Increasing destructiveness of tropical cyclones over the past 30
  years.
\newblock {\em Nature}, 436:686--688, 2005.

\bibitem{Corral_hurricanes}
A.~Corral, A.~Oss\'o, and J.~E. Llebot.
\newblock Scaling of tropical-cyclone dissipation.
\newblock {\em Nature Phys.}, 6:693--696, 2010.

\bibitem{Corral_agu}
A.~Corral and A.~Turiel.
\newblock Variability of {North Atlantic} hurricanes: seasonal versus
  individual-event features.
\newblock In A.~S. Sharma, A.~Bunde, V.~P. Dimri, and D.~N. Baker, editors,
  {\em Extreme Events and Natural Hazards: the Complexity Perspective}, pages
  111--125. Geopress, Washington, 2012.

\bibitem{Peters_np}
O.~Peters and J.~D. Neelin.
\newblock Critical phenomena in atmospheric precipitation.
\newblock {\em Nature Phys.}, 2:393--396, 2006.

\bibitem{Stanley_rmp}
H.~E. Stanley.
\newblock Scaling, universality, and renormalization: {Three} pillars of modern
  critical phenomena.
\newblock {\em Rev. Mod. Phys.}, 71:S358--S366, 1999.

\bibitem{Bak_book}
P.~Bak.
\newblock {\em How Nature Works: The Science of Self-Organized Criticality}.
\newblock Copernicus, New York, 1996.

\bibitem{Zapperi_branching}
S.~Zapperi, K.~B. Lauritsen, and H.~E. Stanley.
\newblock Self-organized branching processes: Mean-field theory for avalanches.
\newblock {\em Phys. Rev. Lett.}, 75:4071--4074, 1995.

\bibitem{Avnir}
D.~Avnir, O.~Biham, D.~Lidar, and O.~Malcai.
\newblock Is the geometry of nature fractal?
\newblock {\em Science}, 279(5347):39--40, 1998.

\bibitem{Leitao}
J.~C. Leit{\~a}o, J.~M. Miotto, M.~Gerlach, and E.~G. Altmann.
\newblock Is this scaling nonlinear?
\newblock {\em R. Soc. Open Sci.}, 3(7):150649, 2016.

\bibitem{Bauke}
H.~Bauke.
\newblock Parameter estimation for power-law distributions by maximum
  likelihood methods.
\newblock {\em Eur. Phys. J. B}, 58:167--173, 2007.

\bibitem{White}
E.~P. White, B.~J. Enquist, and J.~L. Green.
\newblock On estimating the exponent of power-law frequency distributions.
\newblock {\em Ecol.}, 89:905--912, 2008.

\bibitem{Clauset}
A.~Clauset, C.~R. Shalizi, and M.~E.~J. Newman.
\newblock Power-law distributions in empirical data.
\newblock {\em SIAM Rev.}, 51:661--703, 2009.

\bibitem{Corral_nuclear}
A.~Corral, F.~Font, and J.~Camacho.
\newblock Non-characteristic half-lives in radioactive decay.
\newblock {\em Phys. Rev. E}, 83:066103, 2011.

\bibitem{Voitalov_krioukov}
I.~{Voitalov}, P.~{van der Hoorn}, R.~{van der Hofstad}, and D.~{Krioukov}.
\newblock Scale-free networks well done.
\newblock {\em Phys. Rev. Research}, 1:033034, 2019.

\bibitem{Corral_garcia_moloney_font}
A.~Corral, R.~Garcia-Millan, N.~R. Moloney, and F.~Font-Clos.
\newblock Phase transition, scaling of moments, and order-parameter
  distributions in {Brownian} particles and branching processes with
  finite-size effects.
\newblock {\em Phys. Rev. E}, 97:062156, 2018.

\bibitem{Malevergne_Sornette_umpu}
Y.~Malevergne, V.~Pisarenko, and D.~Sornette.
\newblock Testing the {Pareto} against the lognormal distributions with the
  uniformly most powerful unbiased test applied to the distribution of cities.
\newblock {\em Phys. Rev. E}, 83:036111, 2011.

\bibitem{Corral_Arcaute}
A.~Corral, F.~Udina, and E.~Arcaute.
\newblock Truncated lognormal distributions and scaling in the size of
  naturally defined population clusters.
\newblock {\em Phys. Rev. E}, 101:042312, 2020.

\bibitem{Corral_epidemics_pre}
A.~Corral.
\newblock Tail of the distribution of fatalities in epidemics.
\newblock {\em Phys. Rev. E}, 103:022315, 2021.

\bibitem{Serra_Corral_Zipf}
M.~Serra-Peralta, J.~Serr\`a, and A.~Corral.
\newblock Lognormals, power laws and double power laws in the distribution of
  frequencies of harmonic codewords from classical music.
\newblock {\em Sci. Rep.}, 12:2615, 2022.

\bibitem{Corral_Deluca}
A.~Deluca and A.~Corral.
\newblock Fitting and goodness-of-fit test of non-truncated and truncated
  power-law distributions.
\newblock {\em Acta Geophys.}, 61:1351--1394, 2013.

\bibitem{Gerlach_Altmann_prl}
M.~Gerlach and E.~G. Altmann.
\newblock Testing statistical laws in complex systems.
\newblock {\em Phys. Rev. Lett.}, 122:168301, 2019.

\bibitem{Moore_Altmann_PRX}
J.~M. Moore, G.~Yan, and E.~G. Altmann.
\newblock Nonparametric power-law surrogates.
\newblock {\em Phys. Rev. X}, 12:021056, 2022.

\bibitem{Broderick}
T.~Broderick, M.~Dud\'{\i}k, G.~Tkacik, R.~E. Schapireb, and W.~Bialek.
\newblock Faster solutions of the inverse pairwise {Ising} problem.
\newblock {\em arXiv}, 0712.2437, 2007.

\bibitem{Coles}
S.~Coles.
\newblock {\em An Introduction to Statistical Modeling of Extreme Values}.
\newblock Springer, London, 2001.

\bibitem{Mantegna_Stanley}
R.~N. Mantegna and H.~E. Stanley.
\newblock {\em An Introduction to Econophysics}.
\newblock Cambridge Univ. Press, Cambridge, UK, 2000.

\bibitem{Sornette_critical_book}
D.~Sornette.
\newblock {\em Critical Phenomena in Natural Sciences}.
\newblock Springer, Berlin, 2nd edition, 2004.

\bibitem{Mitz}
M.~Mitzenmacher.
\newblock A brief history of generative models for power law and lognormal
  distributions.
\newblock {\em Internet Math.}, 1 (2):226--251, 2004.

\bibitem{Newman_05}
M.~E.~J. Newman.
\newblock Power laws, {Pareto} distributions and {Zipf}'s law.
\newblock {\em Cont. Phys.}, 46:323--351, 2005.

\bibitem{Penland}
C.~Penland and P.~D. Sardeshmukh.
\newblock Alternative interpretations of power-law distributions found in
  nature.
\newblock {\em Chaos}, 22(2):023119, 2012.

\bibitem{Morina_R}
J.~{del Castillo}, I.~Serra, M.~Padilla, and D.~Mori{\~n}a.
\newblock {Fitting Tails by the Empirical Residual Coefficient of Variation:
  The ercv Package}.
\newblock {\em {R Journal}}, 11(2):56--68, 2019.

\bibitem{lin2004stratiform}
J.~Lin, B.~Mapes, M.~Zhang, and M.~Newman.
\newblock Stratiform precipitation, vertical heating profiles, and the
  {Madden--Julian} oscillation.
\newblock {\em J. Atm. Sci.}, 61(3):296--309, 2004.

\bibitem{Vitart2017TheST}
Frederic Vitart, Constantin Ardilouze, A.~Bonet, A.~M. Brookshaw, M.~Chen,
  C.~Codorean, Michel D{\'e}qu{\'e}, Laura Ferranti, Enrico Fucile, Manuel
  Fuentes, Harry~H. Hendon, Jerry Hodgson, Hyun-Suk Kang, Arun~S Kumar, Hai
  Lin, G.~X. Liu, Xin Liu, Piero Malguzzi, Ioannis Mallas, Menelaos~N
  Manoussakis, Daniele Mastrangelo, Craig MacLachlan, Peter McLean, Atsushi
  Minami, Richard Ml{\'a}dek, Tetsuo Nakazawa, Suhair~Khalaf Najm, Yu~Nie,
  Michel Rixen, Andrew Robertson, Paolo~M. Ruti, C.~Sun, Yuhei Takaya,
  Mikhail~A. Tolstykh, Fabio Venuti, Duane~E. Waliser, Steven~James Woolnough,
  Tongwen Wu, D~J Won, H.~Xiao, R.~R. Zaripov, and Liyue Zhang.
\newblock The subseasonal to seasonal {(S2S)} prediction project database.
\newblock {\em Bull. Am. Meteor. Soc.}, 98:163--173, 2017.

\bibitem{ZhangLing17}
Ch. Zhang and J.~Ling.
\newblock Barrier effect of the {Indo-Pacific Maritime Continent} on the {MJO}:
  Perspectives from tracking {MJO} precipitation.
\newblock {\em J. Clim.}, 30(9):3439--3459, 2017.

\bibitem{Randall_book}
D.~Randall.
\newblock {\em Atmosphere. Clouds, and Climate}.
\newblock Princeton Univ. Press, Princeton, New Jersey, 2012.

\bibitem{Zhang_theories_mjo}
C.~Zhang, A.~F. Adames, B.~Khouider, B.~Wang, and D.~Yang.
\newblock Four theories of the {Madden-Julian} oscillation.
\newblock {\em Rev. Geophys.}, 58(3):e2019RG000685, 2020.
\newblock e2019RG000685 2019RG000685.

\bibitem{Hottovy_CAFE}
S.~Hottovy.
\newblock Unified spectrum of tropical rainfall and waves in a simple
  stochastic model.
\newblock {\em CAFE Final Conference, Barcelona}, 2022.

\bibitem{Hand_mjo}
E.~Hand.
\newblock The storm king.
\newblock {\em Science}, 350(6256):22--25, 2015.

\bibitem{wheeler2004all}
M.~C. Wheeler and H.~H. Hendon.
\newblock An all-season real-time multivariate {MJO} index: Development of an
  index for monitoring and prediction.
\newblock {\em Month. Weather Rev.}, 132(8):1917--1932, 2004.

\bibitem{Baro_Corral}
J.~Bar\'o, A.~Corral, X.~Illa, A.~Planes, E.~K.~H. Salje, W.~Schranz, D.~E.
  Soto-Parra, and E.~Vives.
\newblock Statistical similarity between the compression of a porous material
  and earthquakes.
\newblock {\em Phys. Rev. Lett.}, 110:088702, 2013.

\bibitem{mjo_index_data}
{Bureau of Meteorology, Australian Government}.
\newblock Madden-julian oscillation (mjo).
\newblock {\em
  \url{http://www.bom.gov.au/climate/mjo/graphics/rmm.74toRealtime.txt}}, 2022.

\bibitem{Barnes_causal_MJO}
S.~M. Samarasinghe, Ch. Connolly, E.~A. Barnes, I.~Ebert-Uphoff, and L.~Sun.
\newblock Strengthened causal connections between the {MJO} and the {North
  Atlantic} with climate warming.
\newblock {\em Geophys. Res. Lett.}, 48(5):e2020GL091168, 2021.

\bibitem{Matthews08}
Adrian~J. Matthews.
\newblock Primary and successive events in the {Madden–Julian} oscillation.
\newblock {\em Q. J. Roy. Meteor. Soc.}, 134(631):439--453, 2008.

\bibitem{Corral_Deluca_arxiv}
A.~Corral, A.~Deluca, and R.~{Ferrer-i-Cancho}.
\newblock A practical recipe to fit discrete power-law distributions.
\newblock {\em ArXiv}, 1209:1270, 2012.

\bibitem{Corral_Cancho}
A.~Corral, I.~Serra, and R.~{Ferrer-i-Cancho}.
\newblock Distinct flavors of {Zipf's} law and its maximum likelihood fitting:
  Rank-size and size-distribution representations.
\newblock {\em Phys. Rev. E}, 102:052113, 2020.

\bibitem{footnote_mjo2}
This relation has been published in many papers, without a clear original
  souce. {A} relatively old reference (but for sure not the first one) is
  {Ref.} \cite{Corral_thesis}.

\bibitem{Corral_thesis}
A.~Corral.
\newblock {\em Complex behavior in slowly driven dynamical systems: sandpiles,
  earthquakes, biological oscillators}.
\newblock PhD thesis, University of Barcelona, 1997.

\bibitem{Kalbfleisch2}
J.~D. Kalbfleisch and R.~L. Prentice.
\newblock {\em The Statistical Analysis of Failure Time Data}.
\newblock Wiley, Hoboken, NJ, 2nd edition, 2002.

\bibitem{Holme_private}
We are indebted to {Petter Holme} for a {Twitter} communication.

\end{thebibliography}

%\begin{figure}[ht]
%%\includegraphics[width=1.\columnwidth]{figs/fig_Ddur_months_phasein.pdf}\\
%%\includegraphics[width=1.\columnwidth]{figs/fig_Dn_months_phasein.pdf}
%\caption{
%MESES Y FASES INICIALES!!
%}
%\label{}
%\end{figure}

%BORRAR EL SIGUIENTE PARRAFO!!!!
%A scatter plot between $n$ and the event duration $d$ shows that the crossover value
%$d=27$ days could be associated with a value of $n\simeq 8$, 
%which would mean that events with $n=9$ or $n=10$, etc., 
%are much more unlikely that events with 
%$n=7$.

%LO DE ABAJO BORRARLO:{\tiny
%In order to test this, we compare the unrestricted distribution of durations, $f(d)$,
%with the distribution of durations restricted to $n\le 8$, 
%that is, 
%$f(d\,|\, n\le 8)$.
%Figure %\ref{vuelta1} 
%confirms that the tail of the distribution is mostly
%composed by events with $n>8$, 
%whereas the body of the distribution is given by events with $n\le 8$.
%}

%\begin{figure}[ht]
%\includegraphics[width=1.\columnwidth]{figs/1vuelta.pdf}
%\caption{
%Empirical probability density of sizes restricted  to ????,
%compared to the unrestricted distribution.
%}
%\label{vuelta1}
%\end{figure}
%

\end{document}